\RequirePackage[]{silence}
\documentclass[11pt]{article}
\usepackage{fullpage}
\usepackage{color}
\usepackage{graphicx}
\usepackage[framemethod=tikz]{mdframed}
\usepackage{algorithm}
\usepackage[noend]{algpseudocode}
\usepackage{epstopdf}

\usepackage[pagebackref=false]{hyperref}
\hypersetup{
    unicode=false,          
    colorlinks=true,        
    linkcolor=red,          
    citecolor=green,        
    filecolor=magenta,      
    urlcolor=cyan           
}
\usepackage{amsmath,amssymb,amsthm,cite}
\usepackage[capitalize]{cleveref}

\usepackage[textsize=tiny]{todonotes}

\newcommand{\mtodo}[1]{\todo[color=blue!20]{#1}}

\newcommand{\security}{{k}_{S}}

\newcommand{\floor}[1]{\left\lfloor #1 \right\rfloor}
\newcommand{\ceil}[1]{\left\lceil #1 \right\rceil}

\newtheorem{theorem}{Theorem}[section]
\newtheorem{observation}[theorem]{Observation}
\newtheorem{definition}[theorem]{Definition} 
\newtheorem{lemma}[theorem]{Lemma}
\newtheorem{claim}[theorem]{Claim}
\newtheorem{corollary}[theorem]{Corollary}
\newtheorem{remark}[theorem]{Remark}
\newtheorem*{hypothesis}{Invariability Hypothesis of \cite{GHS13} (a simpler version)}

\Crefname{theorem}{Theorem}{Theorems}
\Crefname{lemma}{Lemma}{Lemmas}
\Crefname{claim}{Claim}{Claims}
\Crefname{remark}{Remark}{Remarks}
\Crefname{corollary}{Corollary}{Corollaries}
\Crefname{proposition}{Proposition}{Propositions}
\Crefname{definition}{Definition}{Definitions}
\Crefname{observation}{Observation}{Observations}

\algnewcommand\algorithmicswitch{\textbf{switch}}
\algnewcommand\algorithmiccase{\textbf{case}}

\algdef{SE}[SWITCH]{Switch}{EndSwitch}[1]{\algorithmicswitch\ #1\ \algorithmicdo}{\algorithmicend\ \algorithmicswitch}%
\algdef{SE}[CASE]{Case}{EndCase}[1]{\algorithmiccase\ #1}{\algorithmicend\ \algorithmiccase}%
\algtext*{EndSwitch}%
\algtext*{EndCase}%

\algnewcommand\algorithmicwithprob{\textbf{with probability}}
\algnewcommand\algorithmicotherwise{\textbf{otherwise}}

\algdef{SE}[WithProb]{WithProb}{EndWithProb}[1]{\algorithmicwithprob\ #1\ \algorithmicdo}{\algorithmicend\ \algorithmicwithprob}%
\algdef{Ce}[Otherwise]{WithProb}{Otherwise}{EndWithProb}
  {\algorithmicotherwise}%
\algtext*{EndWithProb}%
\algtext*{EndOtherwise}%
\newcommand{\FullOrShort}{full}
\ifthenelse{\equal{\FullOrShort}{full}}{
	
	  \newcommand{\fullOnly}[1]{#1}
	  \newcommand{\shortOnly}[1]{}
		\renewcommand{\paragraph}[1]{\vspace{0.15cm}\noindent {\bf #1}:}
		
  }{
	  \newcommand{\fullOnly}[1]{}
	  \newcommand{\shortOnly}[1]{#1}
		\renewcommand{\paragraph}[1]{\vspace{0.12cm}\noindent {\bf #1}:}
		\usepackage{enumitem}
		\setitemize{itemsep=3pt, topsep=3pt,parsep=3pt,partopsep=3pt}
  }

\begin{document}

\date{}
\title{Optimal Error Rates for Interactive Coding II: \\ Efficiency and List Decoding}

\author{
Mohsen Ghaffari\\
MIT\\
\texttt{ghaffari@mit.edu}
\and
Bernhard Haeupler\\
Microsoft Research\\
\texttt{haeupler@cs.cmu.edu}}

\maketitle

\newcommand{\listen}{\text{listen}}
\newcommand{\eps}{\epsilon}

\begin{abstract}
We study coding schemes for error correction in interactive communications. Such \emph{interactive coding schemes} simulate any $n$-round interactive protocol using $N$ rounds over an adversarial channel that corrupts up to $\rho N$ transmissions. Important performance measures for a coding scheme are its maximum \emph{tolerable error rate} $\rho$, \emph{communication complexity} $N$, and \emph{computational complexity}.

\smallskip

We give the first coding scheme for the standard setting which performs optimally in all three measures: Our randomized non-adaptive coding scheme has a near-linear computational complexity and tolerates any error rate $\delta < 1/4$ with a linear $N = \Theta(n)$ communication complexity. This improves over prior results \cite{BR11,BK12,BN13,GHS13} which each performed well in two of these measures. 

\smallskip

We also give results for other settings of interest, namely, the first computationally and communication efficient schemes that tolerate $\rho < \frac{2}{7}$ adaptively, $\rho < \frac{1}{3}$ if only one party is required to decode, and $\rho < \frac{1}{2}$ if list decoding is allowed. These are the optimal tolerable error rates for the respective settings. These coding schemes also have near linear computational and communication complexity\footnote{Our boosting technique leads to computationally and communication efficient list decoding schemes. Its exact communication complexity depends on the list decoder that is boosted. In particular, if we start with the simple list decoder from \cite{GHS13}, which has a quadratic communication complexity, we achieve a near linear time coding scheme with communication complexity $N = n 2^{O(\log^* n \, \cdot \, \log{\log^* n})} = o(n \log \log \ldots \log n)$. If we start with the exponential time and linear communication complexity list decoder of \cite{BE14}, a near linear time coding scheme with linear communication complexity $N = O(n)$ is obtained. Since our reductions preserve both communication and computational complexity, these are also the complexities of our unique decoding schemes.}.

\smallskip

These results are obtained via two techniques: We give a \emph{general black-box reduction} which reduces unique decoding, in various settings, to list decoding. We also show how to boost the computational and communication efficiency of any list decoder to become near linear\textcolor{red}{\footnotemark[1]}.

%
%
%
%
%
%
\end{abstract}

\shortOnly{
\setcounter{page}{0}
\thispagestyle{empty}
}
\newpage
\fullOnly{
\begingroup
\hypersetup{linkcolor=blue}

{\small \tableofcontents}
\endgroup
\newpage
}

\section{Introduction}

``Interactive Coding'' or ``Coding for Interactive Communication'' can be viewed as an extension of error correcting codes to interactive communications. Error correcting codes enable a party to communicate a message through a channel to another party even if a constant fraction of the symbols of the transmission are corrupted by the channel. This coding for ``one-way communication" is achieved by adding redundancy to the message, that is, by coding an $n$-bit message into a slightly longer $N$-symbol coded message over some finite alphabet. 
Interactive coding schemes, as introduced by Schulman~\cite{Schulman}, generalize this to two-way interactive communication: they enable two parties to perform their interaction even if a constant fraction of the symbols are corrupted by the channel. This robustness against \emph{errors} is again achieved by adding redundancy to the interactive communication, now by transforming the original interaction protocol $\Pi$ which uses $n$ \emph{communication rounds} into a new coded protocol $\Pi'$ which has longer length, ideally still $N = O(n)$ rounds. Running this coded protocol $\Pi'$ both parties can recover the outcome of $\Pi$ even if a constant fraction $\rho$ of the symbols are corrupted during the execution of $\Pi'$. 

Similar to the classical error correcting codes, important performance measures of an interactive coding scheme are: the \emph{maximum tolerable error-rate} $\rho$ that can be tolerated, the \emph{communication complexity} $N$, and the \emph{computational complexity} of the coding and decoding procedures.

For error correcting codes the classical results of Shannon show that for any constant error-rate below $1/2$, there exist codes with $N = O(n)$, that is, with a constant redundancy factor. Deterministic linear time encoding and decoding procedures that achieve this optimal error rate and redundancy are also known \cite{spielman1996linear}. Interestingly, error correcting codes can also tolerate any constant error rate below $1$ if one relaxes the decoding requirement to \emph{list decoding}~\cite{elias1957listdecoding}, that is, allows the receiver to output a (constant size) list of outputs of which one has to be correct. Computationally efficient list decoders are however a much more recent discovery~\cite{SudanListDecoding,guruswami2001list}.

The interactive coding setting is more involved and less well understood: In 1993 Schulman~\cite{Schulman} gave an interactive coding scheme that tolerates an adversarial error rate of $\rho = 1/240$ with a linear communication complexity $N = O(n)$. In a more recent result that revived this area, Braverman and Rao~\cite{BR11} increased the tolerable error rate to $\rho \leq 1/4-\eps$, for any constant $\eps>0$, and showed this bound to be tight if one assumes the schedule of which party transmits at what round to be fixed ahead of time, that is, if the coding scheme is required to be \emph{non-adaptive}. Both protocols have an \emph{exponential computational complexity}.

More efficient protocols were given in \cite{GMS11, Braverman12,BK12,BN13}: Gelles, Moitra, and Sahai\cite{GMS11} give efficient ranzomized coding schemes for random instead of adversarial errors. The protocol presented by Braverman in\cite{Braverman12} uses sub-exponential time and tolerates an error rate of at most $1/40$. Most related to this paper is the randomized coding scheme of Brakerski and Kalai~\cite{BK12}, which runs in quadratic time and tolerates any error rate below $1/16$, and its extension by Brakerski and Naor~\cite{BN13}, which runs in near-linear time and tolerates some small unspecified constant error rate. These protocols therefore \emph{compromise on the maximum tolerable error-rate} to achieve computational efficiency. 

Our first result shows that, in this standard setting, both computational complexity and an optimal maximum tolerable error-rate are achievable simultaneously:

\begin{theorem}\label{thm:main14}
For any constant $\eps>0$ and $n$-round protocol $\Pi$ there is a randomized non-adaptive coding scheme that robustly simulates $\Pi$ against an adversarial error rate of $\rho \leq \frac{1}{4}-\eps$ using $N = O(n)$ rounds, a near-linear $n \log^{O(1)} n$ computational complexity, and failure probability $2^{-\Theta(n)}$.
\end{theorem}

Protocols without the non-adaptivity restriction and other interactive coding settings of interest were studied in \cite{GHS13,AGS13,FGOS13,BE14}: Particularly, \cite{GHS13, AGS13} study different notions of adaptivity, \cite{FGOS13} studies a setting with shared (hidden) randomness, and the concurrent work of \cite{BE14} investigates list decoding and also tolerable error rate regions for settings with two separate unidirectional channels with different error rates $\alpha,\beta$. Most related to this paper is \cite{GHS13}, which showed that the maximum tolerable error rate can be improved from $1/4$ to $2/7$ by using \emph{adaptive} coding schemes, or to $1/2$ by allowing list decoding. They also showed these bounds on the maximum tolerable error rate to be optimal even if an unbounded amount of communication is allowed. However, the coding schemes achieving these error rates required \emph{polynomially large communication complexity $N = O(n^2)$}. 

We give the first computationally and communication efficient coding schemes that achieve the optimal error-rates in these settings:

\begin{theorem}\label{thm:mainOthers}
For any constant $\eps>0$ and $n$-round protocol $\Pi$, there are the following coding schemes that robustly simulate $\Pi$:
\begin{itemize}
	\item[(A)] An adaptive unique decoding protocol tolerating error-rate $\rho\leq\frac{2}{7}-\eps$.
	\item[(B)] A non-adaptive one-sided unique decoding protocol, in which only one fixed party uniquely decodes, tolerating error-rate $\rho \leq \frac{1}{3}-\eps$.
	\item[(C)] A non-adaptive list decoding protocol with an $O(\frac{1}{\eps^2})$ list size tolerating error-rate $\rho\leq \frac{1}{2}-\eps$.
\end{itemize}
These coding schemes are all randomized, use $N=O(n)$ communication rounds\footnote{A part in achieving these coding schemes is to boost list-decoders. While the boosting will always reduce the computational complexity and communication to near linear the exact communication complexity of the final scheme depends on that of the initial list decoder that is boosted. If we start with the simple list decoder from~\cite{GHS13}, which has a quadratic communication complexity, the final communication complexity becomes $N = n 2^{O(\log^* n \, \cdot \, \log{\log^* n})} = n o(\log \log \ldots \log n)$. If we start with the list decoder of Braverman and Efremenko~\cite{BE14}, which has linear communication complexity, the final communication complexity stays linear, that is, $N=O(n)$, while the computational complexity improves from exponential time to near linear time.}, and near-linear $n \log^{O(1)} n$ computational complexity, and have a failure probability of $2^{-\Omega(n)}$. 
\end{theorem}	
An interesting remaining question is to achieve the above results deterministically. In this paper, we already take a first step in that direction by providing non-uniform deterministic coding schemes:

\begin{remark}\label{rmrk:deterministic} For each of the coding schemes in \Cref{thm:main14,thm:mainOthers}, there exists a (non-uniform) deterministic near linear time variant with the same tolerable error-rate and linear communication complexity. It remains open whether these deterministic schemes can be found efficiently. 
\end{remark}

\subsection{Techniques}
Our results rely on two main technical components, a \emph{reduction} from unique decoding to list decoding, and a \emph{boosting technique} for list-decoders.  Next, we give a brief overview over these components:

\subsubsection{Black-box Reductions from Unique Decoding to List Decoding}\label{overview:reduction}

The reduction technique shows a strong connection between unique-decoding and list-decoding for interactive coding. This technique can be roughly viewed as follows: given a ``good" list-decodable coding scheme, we can construct ``good" unique-decoding coding schemes for various settings in a black-box manner: 
\begin{theorem}\label{thm:ResultsReduction}
Given any non-adaptive list decodable coding scheme for an error rate of $1/2 - \eps$ with constant list size, one can construct unique decoding schemes with optimal error rates for various settings, while preserving the asymptotic round complexity and computational efficiency of the list decoding coding scheme. In particular, one can obtain a non-adaptive coding scheme with error rate $1/4 - O(\eps)$, an adaptive coding scheme with error rate $2/7 - O(\eps)$, and a coding scheme with one-sided decoding and error rate $1/3 - O(\eps)$. 
\end{theorem}
The general idea of the reduction is easier to explain in the non-adaptive setting. Intuitively, we use $O(\frac{1}{\eps})$ repetitions, in each of which we simulate the protocol using the provided list decoder. Thus, in each repetition, each party gets constant many guesses (i.e., decodings) for the correct path. The parties keep all edges of these paths and simultaneous with the simulations of each repetition, they send this accumulated collection of edges to each other using error correcting codes. At the end, each party outputs the path that appeared most frequently in (the decodings of) the received collections. Since we the overall error-rate is always less than what the list-decoder tolerates, in some (early enough) repetition the correct path will be added to the collections. From there on, any repetition corrupted with an error-rate less than what the list-decoder tolerates will reinforce the correct path. This will be such that at the end, a majority-based rule is sufficient for finding the correct path. 

\paragraph{Remark} For the reduction in the non-adaptive setting, it suffices if we start with a list-decoder that tolerates a suboptimal error-rate of $1/4-\eps$ instead of the $1/2-\eps$ stated in \Cref{thm:ResultsReduction}. This allows us to interpret the unique decoder of Braverman and Rao\cite{BR11} as a list decoder (with list size one), boost it to become computationally efficient, and finally transform it back to a unique decoder with the same tolerable error rate. We note the reductions for the adaptive setting and the one-sided decoding setting are much more involved and do not allow for such a slack in the error rate of the initial list decoder. This makes it imperative to start with a list decoder tolerating the optimal $1/2-\eps$ error-rate, at least if one wants unique decoding algorithms with optimal tolerable error rates of $2/7$ and $1/3$. 

\subsubsection{Boosting the Efficiency of List Decoding} Given these reductions, the remaining problem is to construct ``good" list-decodable coding schemes. The main technical element we use in this direction is an approach that allows us to \emph{boost} the performance measures of list-decoders. Particularly, this approach takes a list decoder for any short, poly-logarithmic-round, protocol and transforms it into a list decoder for a long, $n$-round, protocol while (essentially) maintaining the same error rate and list size but significantly improving over the communication and computational complexity and failure probability:

\begin{theorem}[simpler version of \Cref{thm:boosting}]\label{thm:simpleboosting} Suppose there is a list-decodable coding scheme for any $\Theta(\log^2 n)$-round protocol that tolerates an error rate of $\rho$ with failure probability $o(1)$ using $O(R \log^2 n)$ rounds, a list size of $s$ and computational complexity $O(T \log^2 n)$. Then, for any $\eps>0$, there is a list-decodable coding scheme for any $n$-round protocol that tolerates an error rate of $\rho -\eps$ with failure probability $2^{-\Omega(n)}$ using $O(R n)$ rounds, a list size of $O(s/\eps)$ and computational complexity $O(T n \log^{O(1)} n)$.
\end{theorem}
Our boosting is inspired by ideas of Brakerski and Kalai\cite{BK12} and Brakerski and Naor\cite{BN13}, which achieve computationally efficient unique decoders. The general approach is to protect the protocol only in (poly-)logarithmic size blocks, which can be done efficiently, and then use hashing to ensure progress and consistency between blocks. The method in \cite{BK12} sacrifices a factor of $4$ in the tolerable error rate and that of \cite{BN13}, which makes the computation near-linear time, sacrifices an additional unspecified constant factor. See the remark in \cite[Section 3.1]{BN13} for discussions.

Our boosting looses only a small additive $\eps$ in the tolerable error rate. The intuitive reason for this difference is as follows: In \cite{BK12}, one factor of $2$ is lost because of using unique decoding as a primitive for protecting the blocks. This allows the adversary to corrupt the blocks from one party by corrupting each block only half. In our boosting, this is circumvented by using list decoding as the primitive, which also allows us to apply boosting recursively which further lets us have block sizes that are poly-logarihmic instead of logarithmic as in \cite{BK12, BN13}. The second factor of $2$ is lost for trying to keep the pointers of both parties consistent while they are following the correct path. As such, these pointers can only move in constant block size steps and any step in the wrong direction costs two good steps (one back and one forward in the right direction). These (ideally) lock-step moves are conceptually close to the approach of Schulman\cite{Schulman}. Our boosting instead is closer to the approach of Braverman and Rao\cite{BR11}; it continuously adds new blocks and in each step tries to interactively find what the best starting point for an extension is. This interactive search is also protected using the same list decoder. Being able to have poly-logarithmic block sizes, instead of logarithmic, proves important in this interactive search.


%
%

\shortOnly{
\vspace{-10pt}
}
\section*{Organization} 
\shortOnly{\vspace{-8pt}}
The rest of this paper is organized as follows: \Cref{sec:interactivecodingsettings} provides the formal definitions of interactive coding setting and its different variations. As a warm up, in \Cref{sec:efficientadversaries}, we study the simpler setting with a computationally bounded adversary. Then, in Sections \ref{sec:listdecodingreductions} and \ref{sec:boosting} we present our \emph{reduction} and \emph{boosting} results. The boosting in \Cref{sec:boosting} leads to coding schemes with an $\tilde{O}(n^2)$ computational complexity. A more involved boosting which leads to an $\tilde{O}(n)$ computational complexity is presented in \Cref{sec:linearBoost}. 

\section{Interactive Coding Settings}\label{sec:interactivecodingsettings}
In this section, we define the interactive coding setup and summarize the different interactive coding settings considered in \cite{GHS13}. We also define the new one-sided decoding setting which is introduced in this work for the first time. We defer an in-depth discussion of the motivation and results for this new setting to \Cref{app:oneway} and provide here only its definition. 

We mainly adopt the terminology from \cite{GHS13}: An $n$-round {\em interactive protocol} $\Pi$ between two players Alice and Bob is given by two functions $\Pi_A$ and $\Pi_B$. For each {\em round} of communication, these functions map (possibly probabilistically) the history of communication and the player's private input to a decision on whether to listen or transmit, and in the latter case also to a symbol of the {\em communication alphabet} $\Sigma$. For {\em non-adaptive} protocols the decision of a player to listen or transmit deterministically depends only on the round number and ensures that exactly one party transmits in each round. In this case, the {\em channel} delivers the chosen symbol of the transmitting party to the listening party, unless the adversary interferes and alters the symbol arbitrarily. In the adversarial channel model with {\em error rate} $\rho$, the number of such errors is at most $\rho n$. For {\em adaptive} protocols the communicating players are allowed to base their decision on whether to transmit or listen (probabilistically) on the complete communication history (see \cite{GHS13} for an in-length discussion of this model). This can lead to rounds in which both parties transmit or listen simultaneously. In the first case no symbols are delivered while in the latter case the symbols received by the two listening parties are chosen by the adversary, without it being counted as an error. The outcome of a protocol is defined to be the transcript of the interaction. 

\paragraph{Robust Simulation}
A protocol $\Pi'$ is said to {\em robustly simulate} a protocol $\Pi$ for an error rate $\rho$ if the following holds: Given any inputs to $\Pi$, both parties can {\em uniquely decode} the transcript of an error free execution of $\Pi$ on these inputs from the transcript of any execution of $\Pi'$ in which at most a $\rho$ fraction of the transmissions were corrupted. This definition extends easily to {\em $s$-list decoding} by allowing both parties to produce a list of $s$ transcripts that is required to include the correct decoding, i.e., the transcript of $\Pi$. Another natural extension is the \emph{one-sided decoding} in which only one of the two parties is required to decode. For a \emph{one-sided decoding} interactive coding setting we assume that the party to decode is fixed and known a priori. We also consider \emph{randomized protocols} in which both parties have access to independent private randomness which is not known to the adversary\footnote{This requirement is useful to develop our coding schemes but it turns out that all our result can be made to work for the case where the adversary knows the randomness (even in advance). See \Cref{subsub:deterministicBoost}.}. 
We say a randomized protocol robustly simulates a protocol $\Pi$ with \emph{failure probability} $p$ if, for any input and any adversary, the probability that the parties correctly (list) decode is at least $1 - p$. We remark that in all settings the protocol $\Pi'$ typically uses a larger alphabet $\Sigma'$ and a larger number of rounds $N$. Throughout this paper we only consider protocols with constant size alphabets. We furthermore denote with a \emph{coding scheme} any algorithm that given oracle access to $\Pi$ gives oracle access to $\Pi'$. We denote with the \emph{computational complexity} of a coding scheme the number of computation steps performed over $N$ accesses to $\Pi'$ assuming that each oracle access to $\Pi$ is a one step operation. 


\paragraph{Canonical Form} 
A non-adaptive protocol is called \emph{balanced} if in any execution both parties talk equally often. We say a balanced protocol is of \emph{canonical form} if it is over binary alphabet and the two parties take turns sending. Any (non-adaptive) protocol with $m=O(n)$ rounds over an alphabet with size $\sigma=O(1)$ can be transformed to a protocol of canonical form with at most $O(m \log{\sigma}) = O(n)$ rounds which is equivalent when used over an error free channel. We therefore assume that any protocol $\Pi$ to be simulated is an $n$-round protocol of canonical form. To define the protocol $\Pi$, we take a rooted complete binary tree $\mathcal{T}$ of depths $n$. For each node, one of the edges towards children is \emph{preferred}, and these \emph{preferred} edges determine a unique path from the root to a leaf. The set $\mathcal{X}$ of the preferred edges at odd levels is given to Alice as input and the set $\mathcal{Y}$ of the preferred edges at even levels is given to Bob. The output of the protocol is the unique path $\mathcal{P}$, called the \emph{common path}, from the root to a leaf formed by following the preferred edges. The protocol succeeds if both Alice and Bob learn the common path $\mathcal{P}$.
\section{Warm Up: Coding for Computationally Bounded Adversaries}\label{sec:efficientadversaries}
In this section we present coding schemes that work against computationally-bounded adversaries. The material here should be seen as a warm up that provides insights for results in the later sections which apply to the full information theoretic setting, where the adversary is not computationally bounded.

\fullOnly{The organization of this section is as follows: \Cref{sec:whycomputationallybounded} explains why computationally bounded adversaries are a simpler yet still informative setting to look at and \Cref{subsec:boundedAdvSetup} gives the formal model. \Cref{sec:boundedadvintuition} then explains the high level intuitions behind the coding schemes in this section which are contained in \Cref{sec:boundedAdvReductions,sec:boundedAdvBoosting}. In particular, in \Cref{sec:boundedAdvReductions}, we explain how to transform (efficient) list-decoding interactive coding schemes to (efficient) unique-decoding coding schemes with almost the same performance measures for adaptive, non-adaptive, and one-sided decoding settings with a bounded adversary. In \Cref{sec:boundedAdvBoosting}, we present a procedure in the bounded-adversary setting that boosts the communication and compuational efficiency of list-decoders.} 

\fullOnly{
\subsection{Why to look at Computationally Bounded Adversaries} \label{sec:whycomputationallybounded}
In \cite{GHS13}, the authors introduced the Invariablity Hypothesis which states that the error rate of an interactive coding setting does not depend on the communication and computational resources of the coding scheme or the computational resources of the adversary. 
\begin{hypothesis}
Any error rate that is tolerable by a randomized computationally unbounded coding scheme with an unbounded amount of communication and a computationally bounded adversarial channel can also be achieved by a deterministic computationally efficient (polynomial time) coding scheme that requires only constant alphabet size and round complexity of $O(n)$ and works against computationally unbounded adversarial channels. 
\end{hypothesis}
The Invariability Hypothesis is helpful in identifying settings which have the same constraints on the tolerable error-rates but are simpler to understand. Studying coding schemes for these settings helps to develop intuition and insights that can then be used in the design of coding schemes for harder and more general settings. 

Following the Invariability Hypothesis, instead of trying to directly design more efficient protocols for the general settings, we focus on the simpler setting with computationally bounded adversary. This choice is motivated by the following two reasons: (a) All impossibility results in \cite{GHS13} for the optimal tolerable error rates feature computationally very simple adversaries. These results therefore hold also for computationally bounded adversaries. This also makes it seem unlikely that adding computational restrictions to the channel changes in any other way the fundamental barriers and possibilities for efficient coding schemes. (b) The setting with computationally bounded adversary allows us to use the powerful cryptographic tools, particularly \emph{public-key signatures}, which leads to drastic conceptual and technical simplifications. Interestingly, as predicted in \cite{GHS13}, many of the insights we develop in the computationally bounded setting carry over to the standard information theoretic setting where the adversary is not computationally bounded which leads to the results in \Cref{thm:main14,thm:mainOthers}. 
}

\subsection{The Computationally Bounded Adversary Interactive Coding Setting}\label{subsec:boundedAdvSetup}
The computationally bounded adversaries we have in mind are adversaries who are restricted to polynomial time computations. These adversaries still have complete access and control over the channel that is only restricted by the error-rate. As always the adversaries do not know the private randomness or the internal state of Alice and Bob (as long as they do not send this randomness over the channel). 
To be able to use the computational boundedness of the adversary, we consider complexity theoretic hardness assumptions. In particular, we require assumptions that are strong enough to provide public-key signature schemes. We refer to \cite{goldreich2003foundations} for an in-depth definition and explanation of these schemes. \shortOnly{Also, a basic (and slightly simplified) operational definition is presented in \Cref{def:signature} of \Cref{app:efficientadversaries}.}\fullOnly{Here, we just present a basic (and slightly simplified) operational definition.} 

\newcommand{\defSignature}{
\begin{definition}[Unforgeable Public-Key Signature Scheme]
\label{def:signature}
A scheme $(G,S,V)$ consisting of the key generator $G$, and signing and verification algorithms $S$ and $V$ is a public-key signature scheme with security parameter $\security = \log^{1+\Theta(1)} n$ iff $G$ allows to create public/private key pairs $(key_{priv},key_{pub})$ that can be used to efficiently sign any bit-string $m$ of length at least $\security$ using $m' = (m',S(m,key_{priv}))$ such that $m'$ is of length $|m'|<2|m|$ and can be efficiently verified to be the signed version of $m$ by checking $V(m', key_{pub})$. The public key encryption scheme is furthermore unforgeable or secure if no computationally bounded adversary with oracle access to the signing function $S(.,key_{priv})$ can produce a signed version of any message not given to the oracle with more then a negligible $n^{-\omega(1)}$ success probability. 
\end{definition}
}
\fullOnly{\defSignature}

\subsection{High-Level Intuition} \label{sec:boundedadvintuition}
Looking at the setting with computationally bounded adversaries reveals intimate connections between coding schemes for the different unique decoding settings and list decodable interactive communications. In particular, using an unforgeable public-key signature scheme makes it possible to view the errors in an interactive communication as follows: each party is talking to multiple personalities, of which only one is the correct party on the other side of the interactions, while the other personalities are made up by the adversary. Besides having conversations with these personalities, the only task left in order to finally be able to output a unique decoding is for both parties to exchange their public keys. This allows them to identify the correct personalities and pick the right conversation in the end. This robust exchange however is exactly equivalent to the much simpler, non-interactive, two-round \emph{exchange problem} solved in \cite{GHS13}. Put together this leads to conceptually simple reductions from unique decoding to list decoding for the bounded adversary settings. A similar way of using secure public-key signatures as a way of identifying personalities can also be used to boost the efficiency of a list decodable coding scheme. In particular, the parties can cut a conversation in smaller pieces, which are computationally easier to simulate, and then identify which pieces belong together using the signatures. 

\subsection{Reducing Unique Decoding to List Decoding with a Computationally Bounded Adversary}\label{sec:boundedAdvReductions}
In this section we show how to transform a list decodable interactive coding scheme to a coding scheme for unique decoding in various settings. In particular, we consider unique decoding interactive coding schemes for the adaptive and non-adaptive settings as well as their one-sided decoding variants (see \Cref{sec:interactivecodingsettings} for the definitions of these settings). 

\begin{lemma}\label{thm:boundedAdvlistdecodingreductionOneFourth}
In the setting with a computationally bounded adversary, for any constant $\eps>0$, given a balanced (non-adaptive) list decodable coding scheme that tolerates error-rate $1/4 - \eps$, we get a (non-adaptive) unique decodable coding scheme  that tolerates error-rate $1/4 - \eps$ with asymptotically the same round complexity, alphabet size and  computational complexity.
\end{lemma}

\begin{lemma}\label{thm:boundedAdvlistdecodingreductionOneThird}
In the setting with a computationally bounded adversary, for any constant $\eps>0$, given a balanced (non-adaptive) list decodable coding scheme that tolerates error-rate $1/2 - \eps$, we get a (non-adaptive) one-sided unique decodable coding scheme that tolerates error-rate $1/3 - \eps$ with asymptotically the same round complexity, alphabet size and computational complexity.
\end{lemma}

\begin{lemma}\label{thm:boundedAdvlistdecodingreductionTwoSeventh}
In the setting with a computationally bounded adversary, for any constant $\eps>0$, given a balanced (non-adaptive) list decodable coding scheme that tolerates error-rate $1/3 - \eps$, we get an adaptive unique decodable coding scheme that tolerates error-rate $2/7 - \eps$ with asymptotically the same round complexity, alphabet size and computational complexity.
\end{lemma}

As noted in the introduction the error-rates achieved by these reductions are optimal~\cite{GHS13}. The algorithms achieving these reductions are furthermore extremely simple and natural and rely on the same idea: 

\paragraph{Reduction Template} We first change the protocol $\Pi$ to a protocol $\tilde{\Pi}$ which first executes $\Pi$ and after completion has both parties send each other their public keys and a description of the (just computed) common path of $\Pi$, signed with their private key. In the simulation, the list decodable coding scheme is used on $\tilde{\Pi}$ and run until both parties can list decode. For both parties the list of decodings contains the common path of $\Pi$ signed with the private key of the other party, it may also contain other strings made up by the adversary which look like a signed version of a path, albeit signed with a different key (since the adversary cannot forge signatures). Hence, once a party learns the public key of the other party, he/she can pick out the correct decoding in the end. In order to exchange their public keys, Alice and Bob run an Exchange Protocol simultaneous with the above simulation (by slightly increasing the alphabet size). The error-rates tolerable in this way are therefore essentially the same as the ones established in \cite{GHS13} for the Exchange Protocol.

\shortOnly{We here present the proof of \Cref{thm:boundedAdvlistdecodingreductionOneFourth}. The proof of \Cref{thm:boundedAdvlistdecodingreductionOneThird} and \Cref{thm:boundedAdvlistdecodingreductionTwoSeventh} are similar in nature, although more involved, and are deferred to \Cref{app:efficientadversaries}.}

\begin{proof}[Proof of \Cref{thm:boundedAdvlistdecodingreductionOneFourth}]
Both parties run the balanced robust list decodable coding scheme of $\tilde{\Pi}$ which is assumed to tolerate an error-rate of $1/4 - \eps$. At the end of this, both parties can therefore list decode. The parties also encode their public key into an error correcting code of distance $1 - \eps$ and send the resulting codeword symbol by symbol in every transmission they make. With a global error-rate of $1/4 - \eps$ and because both parties talk equally often, the codeword received by each party has an corruption rate of at most $1/2 - 2\eps$. This allows both parties to correctly decode the public key of the other party in the end. Then they pick out the right decoding out of the list. 
\end{proof}

\newcommand{\proofBoundedAdvlistdecodingreductionOneThird}{
\begin{proof}\shortOnly{[Proof of \Cref{thm:boundedAdvlistdecodingreductionOneThird}]}
Both parties run the balanced robust list decodable coding scheme of $\tilde{\Pi}$ for $2/3$ of the time which is assumed to tolerate an error-rate of $1/2 - \eps$. Since the global error-rate is $1/3 - \eps$ the relative error-rate in this part is at most $1/2 - 1.5\eps$ which therefore allows both parties to list decode. In parallel to this Bob again encodes his public key into an error correcting code and sends this encoding symbol by symbol to Alice whenever he sends, including in the last $1/3$ fraction of the protocol in which only Bob sends and Alice listens. This results in Alice listening a $2/3$ fraction of the protocol making the relative error-rate on transmissions from Bob to her also at most $1/2 - 1.5\eps$. If Bob uses an error correcting code with distance larger than $1 - 3 \eps$ then Alice can correctly decode Bob's public key in the end and therefore also pick out the correct decoding from the list decoder. 
\end{proof}
}\fullOnly{\proofBoundedAdvlistdecodingreductionOneThird}

\newcommand{\proofBoundedAdvlistdecodingreductionTwoSeventh}{
\begin{proof}\shortOnly{[Proof of \Cref{thm:boundedAdvlistdecodingreductionTwoSeventh}]}
Both parties run the balanced robust list decoder during the first $6/7$ fraction of the protocol. The relative error-rate in this part is at most $7/6(2/7 - \eps) < 1/3 - \eps$. This means that any list decoder that tolerates an error-rate of $1/3 - \eps$ is sufficient for both parties to be able to list decode. 
In parallel both parties again encode their public key into an error correcting code of distance $1 - \eps$ and send their codeword symbol by symbol using the exchange protocol introduced in \cite{GHS13}. In particular, they send equally often for the first $6/7$ fraction of the time at which point at least of the parties recovers the other parties public key securely. For the last $1/7$ fraction of the simulation, any party that has not safely recovered the other parties key listens while any party who has already decoded sends. This guarantees that in the end both parties know their counterpart's public key and can therefore pick out the right decoding from their list. This is possible assuming an error-rate of at most $2/7 - \eps$. We refer for a proof of this to \cite{GHS13}. 
\end{proof}
}\fullOnly{\proofBoundedAdvlistdecodingreductionTwoSeventh}

\subsection{Boosting List-Decodable Coding Schemes with a Computationally Bounded Adversary}\label{sec:boundedAdvBoosting}
Here we explain how to boost the quality of list decoders using a public-key signature scheme. More precisely, we show that given a list decoder for $\Theta(\security)$-round protocols, where $\security$ is the security parameter of the public-key signature scheme, we can get a list-decoder coding scheme for any $n$-round protocol. The latter will have the same computational complexity overhead as that of the former. For instance, for the parameter setting of $\security = \log^{O(1)} n$, the list decoder from \cite{GHS13} runs in poly-logarithmic time and has quadratic and therefore a poly-logarithmic multiplicative overhead in its communication complexity. Preserving these polylogarithmic overheads this coding scheme can then be boosted to an $\tilde{O}(n)$-time list-decodable coding scheme running in $\tilde{O}(n)$ communication rounds for any $n$-round protocol, a drastic improvement. We note that the approach here can be seen as a significantly simpler version of the boosting for the standard information theoretic setting presented in \Cref{sec:boosting}. 

\begin{lemma}\label{lem:simplelistdecoding}
Consider a setting with a public-key signature scheme with security parameter $\security = \log^{1 + \Theta(1)} n$ which is secure against a computationally bounded adversary. Suppose that for any $(10\security)$-round protocol $\Pi''$ there is a list decodable protocol $\Pi'''$ that robustly simulates $\Pi$ with list size $s$ against an error rate of $\rho$ in $R \security$ rounds and computational complexity $\security^{O(1)}$. Then, for any $\eps >0$ and any $n$-round protocol $\Pi$ there is a computationally efficient list decodable protocol $\Pi'$ that robustly simulates $\Pi$ with list size $s'=\frac{s}{\eps}$ against an error rate of $\rho - \eps$ using $O(\frac{R}{\eps}\, n)$ rounds, the same alphabet size as $\Pi'''$, and near linear computational complexity $\tilde{O}(n)$. 
\end{lemma}
\begin{proof}
The protocol $\Pi'$ makes use of an an unforgeable public-key signature scheme, that is, both parties generate a random public/private key pair and use the public verification key as an identifier uniquely connecting them to all message signed by them. 

The protocol $\Pi'$ runs in $\frac{n}{\eps\security}$ \emph{meta-rounds}, each consisting of $R\,\security$ rounds. Each of these meta-rounds executes a $(10\security)$-round protocol that is protected using the list decodable coding scheme $\Pi'''$. The goal of each meta-round is to compute a $\security$-long part of $\Pi$. In particular, the $n$-round protocol $\Pi$ is split into $\frac{n}{\security}$ \emph{blocks}, each of length $\security$, and each meta-round tries to compute a new block. This is by robustly simulating the following protocol in each meta-round:

In the first meta-round, the first block of $\Pi$ is computed and then both parties take the common path of this block, sign it, and send each other the signed version of this path. This protocol is of length $5\security$ and can therefore be simulated by a protocol $\Pi'''_1$ in the first meta round. 

For any meta-round $i>1$ both parties first look at the list decodings of the prior meta-rounds to determine which path to continue. The decoding for a meta-round $j<i$ comes with a list of size $s$ of the decodings, and it includes a signed version of (a prefix of) the correct path if the error rate was smaller than $\rho$ in meta-round $j$. Both parties discard decodings that are clearly not such a signed path. They then run a \emph{verification phase} to figure out which paths were sent by the other party. For this both parties simply exchange their public-keys. They then can pick all the correctly signed paths and filter out any paths that was not signed by the other party. Given that the adversary cannot produce a signed version of an incorrect path, it is clear that the remaining paths assembled together result in a correct subpath of the common path $\mathcal{P}$ of $\Pi$. In a short \emph{intersection phase} the parties determine how far they both know this common path by simply exchanging the height or the number of blocks their path contains. After taking the minimum height, the parties have determined a good common starting point in $\Pi$ to continue the computation from there. In an \emph{extension phase} both parties compute the next block of $\Pi$ and again send each other a signed version of this path. This concludes the protocol which is simulated in each meta-round. Since what is actually performed in a meta-round is a robust simulation $\Pi'''_i$ of this protocol, the algorithm makes one block progress along the common path if the error rate in this meta-round is below $\rho$. 

At the end of the protocol $\Pi''$ both parties decode all meta-rounds, group together all decodings of paths belonging to the same public-key and output any complete path in $\Pi$, that is, any path of length $n$ that goes from the root to a leaf. We remark that it is easy to implement this protocol in $O(n \security^{O(1)}) = \tilde{O}(n)$ computational time by keeping the decoding of each meta-round grouped together by their signatures and extending the resulting paths a as new decodings get computed. Therefore, what remains is to prove correctness, that is, to show that at most $s' = s/\eps$ paths are output and that the correct path is among them if the overall error rate was below $\rho - \eps$.

To see that at most $s/\eps$ paths are output it suffices to count the total number of edges decoded during the algorithm: There are $\frac{n}{\eps\security}$ meta-rounds each generating at most $s$ paths of length $\security$ for a total of $\frac{ns}{\eps}$ edges, which can form at most $\frac{s}{\eps}$ paths of length $n$. Note that each edge can count only in the path of one personality, as it is signed with the related signature. To see that the correct path is among the decodings of $\Pi'$, note that any meta-round that did not make progress has to have a $\rho$ fraction of errors. Since the total error rate is at most $\rho - \eps$, at least an $\eps$ fraction of the $\frac{n}{\eps\security}$ meta-rounds make progress. This guaranteed progress of $\frac{n}{\security}$ blocks, or equivalently $n$ edges, implies that the common path of $\Pi$ is indeed complete at the end of the protocol. 
\end{proof}

\section{Reducing Unique Decoding to List Decoding}\label{sec:listdecodingreductions}
For the rest of this paper, we consider the standard information theoretic setting in which the adversary is computationally unbounded. 

In this section, we show how to use a list decodable interactive coding scheme to build equally-efficient coding schemes for adaptive or non-adaptive unique decoding and also one-sided unique decoding. The results in this section can be viewed as counterparts of those of \Cref{sec:boundedAdvReductions}. We first state our results in \Cref{sec:reductionresults}, then provide the protocols achieving them in \Cref{sec:reductionschemes}, and lastly give their analysis in \Cref{sec:reductionproofs}.

\subsection{Results}\label{sec:reductionresults}
We start with the non-adaptive unique-decoding, which is the more standard setting:

\begin{theorem}\label{thm:listdecodingreduction14}
For any constant $\eps>0$, given a balanced list decodable coding scheme with constant list size that tolerates error-rate $1/4 - \eps$, we get a (non-adaptive) unique decodable coding scheme that tolerates error-rate $1/4 - 2\eps$ with asymptotically the same round complexity, alphabet size and  computational complexity.
\end{theorem}

\Cref{thm:listdecodingreduction14} is interesting because of two somewhat curious aspects: (a) As list decoding is a strictly weaker guarantee than unique decoding, this theorem shows that one can get the uniqueness of the decoding essentially for free in the non-adaptive setting. (b) This theorem takes a list decoder that tolerates a suboptimal error-rate---as it is known that list decoders can tolerate error-rate $1/2 - \eps$---and generates a non-adaptive unique decoder which tolerates an optimal error-rate.

Next, we present the reduction for the adaptive unique-decoding setting: 
\begin{theorem}\label{thm:listdecodingreduction27}
For any constant $\eps>0$, given a balanced list decodable coding scheme with constant list size that tolerates error-rate $1/2 - \eps$, we get an adaptive unique decodable coding scheme that tolerates error-rate $2/7 - 2\eps$ with asymptotically the same round complexity, alphabet size and  computational complexity.
%
\end{theorem}

We next present the reduction for the newly introduced setting of one-sided unique decoding, where only one party---which is determined a priori--- has to uniquely decode:
\begin{theorem}\label{thm:listdecodingreduction13}
For any constant $\eps>0$, given a balanced list decodable coding scheme with constant list size that tolerates error-rate $1/2 - \eps$, we get a (non-adaptive) one-sided unique decodable coding scheme that tolerates error-rate $1/3 - 2\eps$ with asymptotically the same round complexity, alphabet size and  computational complexity.
%
\end{theorem}
Note that the $1/3-\eps$ error-rate that can be tolerated by \Cref{thm:listdecodingreduction13} is larger than the $2/7$ error rate of the more standard two-sided setting (or $1/4$ if protocols are not allowed to be adaptive), in which both parties have to decode uniquely. This means that this slightly weaker decoding requirement, which might be all that is needed in some applications, allows for a higher error tolerance. This makes one-sidedness a useful distinction. We also remark that the $1/3$ tolerable error rate is optimal (see \Cref{sec:onesidedLB}). 

\medskip
Lastly, we also explain that using the same techniques, we can reduce the list size of the list decodable coding schemes to $O(1/\eps^2)$: 

\begin{theorem}\label{thm:listdecoding-listsizereduction}
For any constant $\eps>0$, given a balanced list decodable coding scheme for $n$-round protocols that tolerates error-rate $1/2-\eps$ with list size $s$ and round complexity $N'=\Omega(n s/\eps)$, we get a balanced list decodable coding scheme that tolerates error-rate $1/2 - 2\eps$, with constant list size $s'=O(\frac{1}{\eps^2})$ and round complexity $O(N')$, while having asymptotically the same alphabet size and computational complexity.
\end{theorem}

\subsection{Coding Schemes}\label{sec:reductionschemes}

In this section we describe our coding scheme which underlie the reductions stated in Theorems \ref{thm:listdecodingreduction14}, \ref{thm:listdecodingreduction27}, and \ref{thm:listdecodingreduction13}. All three reductions are built in a very similar manner and can be viewed as following a common template. We describe this coding scheme template in \Cref{sec:reductiontemplate} and then give the concrete instantiations for each of the coding schemes in \Cref{sec:reductions}.

\subsubsection{The Template of the Reductions}\label{sec:reductiontemplate}

\paragraph{Parameters and Structure}
We denote the $n$-round protocol to be simulated with $\Pi$, and we assume that $\Pi$ is in the canonical form\footnote{Note that as stated in \Cref{sec:interactivecodingsettings}, any $n$-round protocol with constant alphabet size can be transformed into a protocol in the canonical form with $O(n)$ rounds.}. We further denote with $\Pi'$ the balanced list decoder coding scheme that we assume to exist, which robustly simulates $\Pi$ and we use $N'$, $\rho'$ and $\Sigma'$ to respectively denote the number of rounds, the tolerable error-rate, and the alphabet of $\Pi'$. As $\Pi'$ is balanced, each party transmits for $N'/2$ rounds during $\Pi'$. We denote with $\Pi''$ the new coding scheme that achieves unique decoding and we use $N''$, $\rho''$ and $\Sigma''$ to respectively denote the number of rounds, the tolerable error-rate, and the alphabet of $\Pi'$. 

We partition the $N''$ rounds of $\Pi''$ into two parts: The first part consists of $b_1 N'$ rounds, which are grouped into $b_1$ \emph{blocks} of $N'$ rounds each. This part is called the \emph{joint part} and the blocks are called \emph{joint blocks}. The second part consists of $b_2 N' / 2$ rounds, which are grouped into $b_2$ blocks, consisting of $N'/2$ rounds each. This part is called the \emph{exclusive part} and the blocks in it are called \emph{exclusive blocks}. During the joint blocks, Alice and Bob send equally often. In the exclusive part, only one party is supposed to talk. Which party talks in the exclusive part is either agreed upon in advance (as for \Cref{thm:listdecodingreduction13}) or decided adaptively (as for \Cref{thm:listdecodingreduction27}).

%


\paragraph{Encoding}
During the protocol $\Pi''$, Alice and Bob respectively maintain sets $E_A\subset \mathcal{X}$ and $E_B\subset \mathcal{Y}$, which are subsets of their \emph{preferred edges} (see the \emph{the canonical form} paragraph in \Cref{sec:interactivecodingsettings} for the definitions). In the beginning of the simulation $\Pi''$, these edge-sets are initialized to be empty. Intuitively, these edge-sets correspond to the set of edges Alice and Bob believe could be on their \emph{common path}. 

In each joint block Alice and Bob run the list-decodable simulation $\Pi''$ and obtain a list of $s$ potentially correct common paths. Each party first discards obviously incorrect paths from its list (those containing non-preferred edges owned by themselves) and then adds all owned edges from all remaining paths to its respective edge-set $E_A$ or $E_B$. The size of these edge-sets increases therefore by at most $s n$ edges per block for a total of at most $b_1 s n$ edges. The edges furthermore form a tree, that is, for every edge all the ancestor edges owned by the same party are included as well. This allows one to encode\footnote{This encoding can be also performed with a simple computation by following the Depth First Search walk over the edge-set, encoding left-downwards, right-downwards and upwards moves respectively as $00$, $01$, and $11$.} each such edge-set using $4(b_1 s n)$ bits, because the number of size $b_1 s n$ subtrees of the complete binary tree is at most $2^{4(b_1 s n)}$. 

In addition to running the list decoder in each block and adding edges to the sets $E_A$ and $E_B$ (which we refer to as $E$-sets), both parties also send their current $E$-sets to each other using error correcting codes. At the beginning of each block, both parties encode their current $E$-set into a codeword consisting of $N'/2$ symbols from an alphabet of size $\sigma_{ECC}=O(1/\eps)$ 
using an error correcting code 
with relative distance of $1 - \eps$. This is where the assumption of $N'=\Omega(ns/\eps)$ comes in, as then $N'/2$ is large enough that can contain an error-correcting coded version of messages of length $4b_1ns$ with relative distance $1-\eps$. During the block they add this codeword symbol by symbol to their transmission leading to the output alphabet size being $[\sigma_{ECC}] \times \Sigma'$. In the exclusive part, the party that is speaking uses the $N'/2$ rounds of each block to send the codeword of its (final) $E$-set symbol by symbol over the same output alphabet. 

\paragraph{Decoding}
%
All decoding decisions only rely on the received possibly-corrupted codewords. We describe how the decoding works for Alice; Bob's decoding procedure is the same. For every $i$, Alice combines the symbols received from Bob during block $i$ to the string $x_i$. Without any channel corruptions $x_i$ would correspond to the codeword encoding the set $E_B$ at the beginning of block $i$. Alice decodes $x_i$ to the closest codeword $\hat{x_i}$ which corresponds to the edge-set $\hat{E_i}$ and assigns this decoding a \emph{confidence} $c_i$ is defined as $c_i = 1 - \frac{2\Delta(x_i,\hat{x_i})}{N'/2}$, where $N'/2$ is the length of error-correcting code. Alice then combines $\hat{E_i}$ with all preferred edges she owns and determines whether these edges together give a unique path. If so Alice calls this path $\hat{\tau_i}$ and otherwise she sets $\hat{\tau_i} = \emptyset$. Given a set of decoded paths $\hat{\tau_1},\hat{\tau_2},\ldots$ and their confidences $c_1,c_2,\ldots$ we denote for any path $\tau$ its confidence with $c(\tau) = \sum_{i : \hat{\tau_i} = \tau} c_i$ and define the majority path $\tau_{\max}$ to be the non-empty path that maximizes this confidence. Lastly we define the combined confidence $C$ as $C = \sum_i c_i$.

For \Cref{thm:listdecoding-listsizereduction}, the decoding procedure is slightly different: In each block, Alice list-decodes $x_i$ to the $L=O(1/\eps)$ closest codewords $\hat{x}^1_i$, \dots, $\hat{x}^L_i$ which  respectively correspond to edge-sets $\hat{E}^1_i$, \dots, $\hat{E}^L_i$, and thus paths $\hat{\tau}^1_i$, \dots, $\hat{\tau}^L_i$. All these paths are output in the end of the algorithm. 


\begin{figure}[t]
	\centering
		\includegraphics[width=0.99\textwidth]{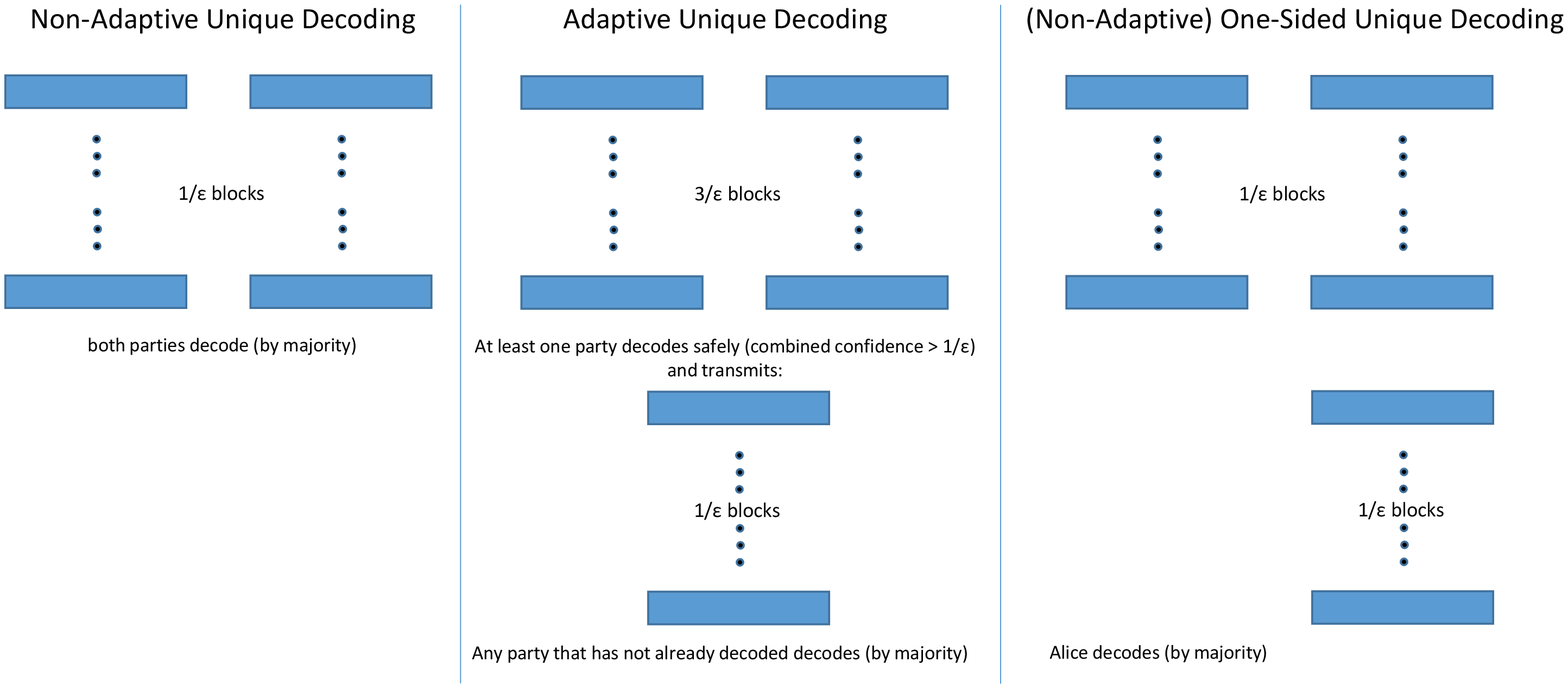}
	\caption{The instantiations of the coding scheme template from \Cref{sec:reductiontemplate} for the three settings considered in \Cref{sec:reductionresults}, namely \Cref{thm:listdecodingreduction14}, \Cref{thm:listdecodingreduction27}, \Cref{thm:listdecodingreduction13}. The instantiation for  \Cref{thm:listdecoding-listsizereduction} is similar to that of \Cref{thm:listdecodingreduction14}.}
	\label{fig:reductions}
\end{figure}

\subsubsection{Setting the Parameters for the Reductions}\label{sec:reductions}
We now describe with what parameters the template from \Cref{sec:reductiontemplate} is employed to lead to the three coding scheme claimed in \Cref{sec:reductionresults}:

For \Cref{thm:listdecodingreduction14}, we use a very simple version of the template from \Cref{sec:reductiontemplate}, in particular, we do not use the exclusive part. The complete protocol consists $N'' = \frac{1}{\eps} N'$ rounds, all in the joint part, which are grouped into $b_1 = \frac{1}{\eps}$ joint blocks. To decode both parties simply output the majority path in the end. 

For \Cref{thm:listdecodingreduction27}, we use the template from \Cref{sec:reductiontemplate} with $N'' = \frac{3.5}{\eps} N'$ rounds which are grouped into $b_1 = \frac{3}{\eps}$ joint blocks and $b_2 = \frac{1}{\eps}$ exclusive blocks. After the joint part both parties determine the quantity $C'' = 2(c(\tau_{\max}) + c(\emptyset)) - C$ and declare the majority path a safe decoding if $C' > 1/\eps$. For the exclusive part both parties make the following adaptive decision: If a party has safely decoded it will transmit in the exclusive part otherwise it will listen and declare the majority path as a decoding in the end. 

For \Cref{thm:listdecodingreduction13}, we use the template from \Cref{sec:reductiontemplate} with $N'' = \frac{1.5}{\eps} N'$ rounds which are grouped into $b_1 = \frac{1}{\eps}$ joint blocks and $b_2 = \frac{1}{\eps}$ exclusive blocks. Assuming that Alice is the party which is interested in decoding in the end, during the exclusive blocks Alice will be listening while Bob is transmitting. To decode Alice outputs the majority path in the end.

For \Cref{thm:listdecoding-listsizereduction}, we use the template from \Cref{sec:reductiontemplate} with no exclusive part: $N'' = \frac{1}{\eps} N'$ rounds, all in the joint part, which are grouped into $b_1 = \frac{1}{\eps}$ joint blocks. At the end, each party outputs all the paths created in the decoding procedure, which are $s'=O(1/\eps^2)$ many.

\subsection{Analyses}\label{sec:reductionproofs}
\shortOnly{As a sample, here we present the analysis for the non-adaptive setting, i.e., \Cref{thm:listdecodingreduction14}. The proofs of the other reductions are deferred to \Cref{app:reduction}. In particular, the analysis of the adaptive setting is considerably more involved.}

%
%
%
%

\fullOnly{
\subsubsection{Analysis for the Non-Adaptive Setting}
}
%
\begin{proof}[Proof of \Cref{thm:listdecodingreduction14}]
As described in \Cref{sec:reductions} we use the template from \Cref{sec:reductiontemplate} with $b_1 = 1/\eps$ joint blocks. 
We show that for the correct path $\tau$ and for both parties the inequality $c(\tau) > (C - c(\tau) - c(\emptyset))$ holds. This means that for both parties the confidence into the path $\tau$ is larger than the confidence in all other (non-emtpy) paths combined. This also implies that the majority path $\tau_{\max}$ is a correct decoding for both parties. 

To prove this we fix one party, say Alice, consider the quantity $C' = c(\tau) - (C - c(\tau) - c(\emptyset))$ for her, and analyze the contribution of each block towards this quantity. We split the execution into two parts according to the first block in which the list decoder succeeded and prove the claim for both parts separately. In particular we define $i^*$ to be the first block at which the list decoder succeeded, that is, the first block after which the edge set $E_A \cup E_B$ contains the common path $\mathcal{P}$ of $\Pi$. We claim that the contribution towards $C'$ of any block $i \neq i^*$ is at least $1 - 4 e_i - 4\eps$ where $e_i$ is the fraction of transmissions with an error in block $i$. 

We first prove this claim for block $i>i^*$. In these blocks the codeword transmitted by Bob corresponds to the correct path $\tau$. Since the error correcting code employed has a distance of at least $1 - \eps$ we get that Alice correctly decodes to $\tau$ if less than a $1/2 - \eps/2$ fraction of Bob's transmissions are corrupted. The confidence $c_i$ of this block then contributes positively to $C'$. It furthermore holds that $c_i = 1 - 2 e_{A}$ where $e_A$ is fraction of errors on Alice which is at most twice the fraction of errors $e_i$ in this block. This makes the contribution of block $i$ at least $1 - 4 e_{i} > 1 - 4 e_i - 4\eps$ as desired. For the case that more than a $1/2 - \eps/2$ fraction of the transmissions to Alice are corrupted she might decode to a different path which makes the confidence of this block contribute negatively to $C'$. We still get that the contribution $-c_i$ is at least $-(1 - 2 (1 - \eps - e_{A})) > 1 - 2\eps - 4e_i$. 

We also need to show that our claim holds for blocks $i<i^*$. The analysis for these blocks is the same except that the codeword sent out by Bob might correspond to the empty path. If few transmissions towards Alice get corrupted she might decode to $\hat{\tau_i} = \emptyset$ which leads to a zero contribution towards $C'$. We need to verify that in this case we do not claim a positive contribution. What saves us is that for any block $i<i^*$ it holds that $e_i > 1/4 - \eps$ since otherwise the list decoder would have been successful in block $i$. This means our assumed contribution of $1 -  4 e_i - 4\eps$ is never larger than zero which completes the proof of the claim for all blocks $i \neq i^*$.

Finally, using the claim, the assumption that the global fraction of errors $e_{ave}$ is at most $\frac{1}{4} - 2\eps$, and summing over the contributions of all blocks we get:
\begin{align*}
C' &\geq \ \ \sum_{i \neq i^*} (1 \ -\ \,4 \ \ e_i \ \ \ \ \ \ - \ \ \frac{4}{\eps}) \ \ \,- \ |c(i^*)|\\
   &> (b_1 - 1) \,- \ 4 \sum_{i} e_i \ \ \ - \ \frac{4 b_1}{\eps} \ \ - \ \ 1\\
	 &= (\frac{1}{\eps} - 1) \ - \ 4 \ b_1 \ e_{ave} \ - \ \ \ 4 \ \ \ \,- \ \ 1\\
	 &\geq \ \frac{1}{\eps} \ \ \ \ \ \ \ \,- \ \ \frac{4}{\eps} \ \left(\frac{1}{4} - 2\eps\right) \ \ \ \ \ \ \,- \ \ 6 \ \ \ = \ \ 2 \ \ > \ \ 0.
\end{align*}
As desired, this shows that an error rate of at most $\frac{1}{4} - 2\eps$ results in both parties recovering the correct path $\tau$ by choosing the majority path. 
\end{proof}

\fullOnly{
\subsubsection{Analysis for the Adaptive Setting}
\begin{proof}[Proof of \Cref{thm:listdecodingreduction27}]
As described in \Cref{sec:reductions} we use the template from \Cref{sec:reductiontemplate} with $N'' = \frac{3.5}{\eps} N'$ rounds grouped into $b_1 = 3/\eps$ joint blocks and $b_2 = 1/\eps$ exclusive blocks. We again choose $\eps_{ECC} < \eps$. 

The key quantity considered in this protocol is $C'' = c(\tau_{\max}) - (C - c(\tau_{\max}) - c(\emptyset)) + c(\emptyset)$. In particular, $C''$ is used by both parties to determine when to decode and whether to transmit in the exclusive blocks. In particular, if the quantity $C''$ for Alice (also denote with $C''_A$) is at least than $1/\eps$ after the joint part than Alice declares the majority path $\tau_{\max}$ her decoding and transmits in the exclusive part. If on the other hand $C''_A < 1 / \eps$ then Alice listens in the exclusive part and decodes to the majority path in the end. Bob bases his decoding and listen/transmit decision on $C''_B$ in the same manner. 

We note that $C''$, similar to the quantity $C'$ used in the proof of \Cref{thm:listdecodingreduction14} and \Cref{thm:listdecodingreduction13}, measures the confidence into the majority path by subtracting the confidence of all other non-empty paths. What is different is that for $C''$ we add the confidence of the empty path in the end. This results in the confidence for an empty path being treated exactly the same as any confidence towards the majority path. This might seem counter intuitive especially since it is easy to see that in the joint part it is possible for the adversary to (i) generate a large amount of confidence for the empty path for one party or (ii) generate a large amout of confidence for one party into an incorrect path making this path the majority path. We will show that really (i) and (ii) cannot hold simultaneously for one party. In particular we will show next that any party for which $C''$ is larger than the decoding threshold of $1/\eps$ decodes correctly when choosing the majority path after the joint part:

\begin{claim}\label{claim:correctSafeMajority}
The majority path of any party that has $C'' \geq 1/\eps$ at any block $t$ after the joint part is correct, that is, is equal to the common path $\tau$ of $\Pi$.
\end{claim}
For sake of contradiction we suppose that Alice has an incorrect majority path and $C''_A > 1/\eps$ after having listened to $t \geq b_1$ blocks in which Bob transmitted. It is easy to see that the quantity $C''(\tau') = c(\tau') - (C - c(\tau') - c(\emptyset)) + c(\emptyset)$ attains its maximum for $\tau' = \tau_{\max}$ so the assumption can also be stated as: After block $t$ there is a path $\tau' \neq \tau$ for which Alice has a combined confidence of $C''(\tau') = c(\tau') - (C - c(\tau') - c(\emptyset)) + c(\emptyset) \geq 1/\eps$.

As in the proof of \Cref{thm:listdecodingreduction14} we analyze the contribution to $C''(\tau')$ in two parts. In particular we again split the execution into two phases according to the first block $i^*$ in which the list decoder succeeded for the first time and prove the claim for both phases separately. The confidence contributed by the first $i^*$ blocks towards $C''(\tau')$ is at most $i^*$ while the number of errors in this phase must be at least $(i^* - 1)N' \cdot (\frac{1}{2} - \eps)$ since otherwise the list decoder would have succeeded before $i^*$. In the remaining $t - i^*$ blocks Bob transmits codewords which decode to the correct path $\tau$ and if uncorrupted would therefore contribute a $-1$ towards $C''(\tau')$ if uncorrupted. On the other hand $(1-\eps)N'/2$ corruptions are enough to turn such a codeword into a codeword for $\tau'$ or an ''empty'' path which contributes a $+1$ to $C''(\tau')$. More generally it is easy to see that the contribution a block $i > i^*$ towards $C''$ is at most $\frac{2E_i}{(1-\eps)N'/2} - 1$ if $E_i$ corruptions happened during this block. 

Adding the contributions of both phases together gives:

\begin{align*}
C''(\tau') &\leq i^* + \sum_{i>i^*}^{t} \left(\frac{2E_i}{(1-\eps)N'/2} - 1\right)\\
           &\leq i^* - (b_1 - i^*) + \frac{4}{(1-\eps)N'}\sum_{i>i^*} E_i \\
					 &\leq - \frac{3}{\eps} + 2 i^* + \frac{4}{(1-\eps)N'}\sum_{i>i^*} E_i 
\end{align*}

Assuming that the total error rate over the $N'' = \frac{3.5}{\eps}N'$ rounds is at most $\frac{2}{7} - 2\eps$ we get that the total number of errors is at most $\frac{3.5}{\eps}N' \cdot (\frac{2}{7} - 2\eps)$. Subtracting the errors from the first phase and observing that 
$$(i^* - 1)N' \cdot \left(\frac{1}{2} - \eps\right) < \frac{i^*}{2}N' - \frac{1}{2}N' - 3.5 N' = \left(\frac{i^*}{2}- 4\right)N'$$ results in: 

\begin{align*}
\sum_{i>i^*} E_i &\leq \frac{3.5}{\eps}N' \cdot \left(\frac{2}{7} - 2\eps\right) - \left(\frac{i^*}{2} - 4\right)N'\\
                 &= \left(\frac{1}{\eps} - 7 - \frac{i^*}{2} + 4\right)N'\\
                 &= \left(\frac{1}{\eps} - \frac{i^*}{2} - 3\right)N'
\end{align*}

Combining these inequalities results in the following contradiction to $C''(\tau')$ being at least $1/\eps$:

\begin{align*}
C''(\tau') &< - \frac{3}{\eps} + 2 i^* + \frac{4}{(1-\eps)N'} \left(\frac{1}{\eps} - \frac{i^*}{2} - 3\right)N'\\
					 &= - \frac{3}{\eps} + 2 i^* + \ \frac{1}{1 - \eps} \ \ \ \ \ \left(\frac{4}{\eps} - 2i^* - 12\right)\\
					 &< - \frac{3}{\eps} + 2 i^* + \frac{4}{\eps} + 8 - 2i^* - 12\\
           &< \ \ \frac{1}{\eps} 
\end{align*}

This proves \Cref{claim:correctSafeMajority} and shows that any party which decodes after the joint phase decodes correctly.

The next claim shows that at least one party will be able to achieve such an early decision:

\begin{claim}\label{claim:onedecoder}
At least one party will decode after the joint part, i.e., either $C''_A$ or $C''_B$ is at least $\frac{1}{\eps}$ after the joint part.
\end{claim}

We will prove \Cref{claim:onedecoder} by showing that after the joint part $C''_
A + C''_B \geq \frac{2}{\eps}$ holds. Actually, we show the slightly stronger statement that the combined confidence in the correct path $\tau$, that is, the confidence quantity $C''(\tau)$ summed up over both parties, is at least $\frac{2}{\eps}$. This is where our definition of $C''$ is helpful. In particular, given that blocks that decode to the empty path increase $C''$ in the same way as blocks decoding to $\tau$ it is not important for the correctness of \Cref{claim:onedecoder} whether the list decoder has succeeded and whether Alice or Bob actually send something that helps to identify the correct path\footnote{This is a crucial difference between $C''$ and the more intuitive quantity $C'$ used in the proofs of \Cref{thm:listdecodingreduction14} and \Cref{thm:listdecodingreduction13}. In particular, the adversary can make $C'_A(\tau) + C'_B(\tau)$ almost as small as zero, e.g., by corrupting all transmissions of Alice during the first $(2-2\eps)/\eps$ blocks leading to a confidence of $(2-2\eps)/\eps$ for the emtpy path for Bob and $(2-2\eps)/\eps$ confidence for an incorrect path for Alice.}. With out any corruptions the $\frac{3}{\eps}$ blocks of the joint part would result in a confidence of $\frac{3}{\eps}$ for each of the parties for a total confidence of $\frac{6}{\eps}$. Roughly speaking the adversary can (fractionally) corrupt only slightly less than the equivalent of $\frac{2}{\eps}$ blocks. This turns less than $\frac{2}{\eps}$ blocks with positive contribution to the combined $C''(\tau)$ into blocks with negative contribution leaving still a total of $\frac{6}{\eps} - 2\frac{2}{\eps} = \frac{2}{\eps}$. More precisely, the contribution of each block to the combined $C''(\tau)$ is at least $2(1 - 2 \frac{e_i}{(1 - \eps)N'})$ where $e_i$ is the number of corruptions in this block. Furthermore, since the total error rate is at most $\frac{2}{7} - 2\eps$, we get that $$\sum_i e_i \leq \frac{3.5}{\eps}N' \cdot (\frac{2}{7} - 2\eps) < \frac{1 - \eps}{\eps}N'.$$
Putting these two inequalities together and summing over all $\frac{3}{\eps}$ blocks shows that the combined confidence is at least $$\frac{6}{\eps} - 4 \frac{\sum_i e_i}{(1 - \eps)N'} > \frac{6}{\eps} - \frac{4}{\eps} = \frac{2}{\eps}.$$
This completes the proof of \Cref{claim:onedecoder}.

\Cref{claim:correctSafeMajority} guarantees that any party who decodes is correct while \Cref{claim:onedecoder} shows that at least one party decodes after the joint part. This implies that if both parties decode after the joint part we are done. If, on the other hand, only one party decodes after the joint part this party transmits in the exclusive blocks while the other party listens. What is left to show is that this party has a correct majority path in the end:

Without loss of generality we assume Alice has not decoded after the joint part. Since the error rate in the joint part is at most $\frac{3.5}{3} (2/7 - 2\eps) < 1/2 - \eps$ it is clear that the list decoder will have succeeded before the exclusive part begins. To show that Alice decodes correctly when she chooses the majority path we follow the arguments given in the proof of \Cref{thm:listdecodingreduction14}. In particular we again consider the quantity $C' = c(\tau) - (C - c(\tau) - c(\emptyset))$ for her and show that $C' > 0$ which can only be true if $\tau_{\max} = \tau$. We also reuse the analysis given in the proof of \Cref{thm:listdecodingreduction14} which shows that the contribution towards Alice's $C'$ of all but one block is at least $1 - 2 e_i - 4\eps$ where $e_i$ is the fraction of transmissions from Bob to Alice with an error in block $i$. Summing over all $\frac{4}{\eps}$ blocks Alice listens to we have $C' \geq \frac{4}{\eps} - 2 \sum_i e_i - 16 - 1.$
Assuming that the global error rate is at most $\frac{2}{7} - 2\eps$ we also get
$$\sum_i e_i \frac{N'}{2} \leq (\frac{2}{7} - 2\eps) \frac{3.5}{\eps}N'$$
from which follows

 $$2 \sum_i e_i \leq 4 (\frac{2}{7} - 2\eps) \frac{3.5}{\eps} \leq \frac{4}{\eps} - 28.$$
This shows that $C' > 0$ and implies that the majority path which Alice chooses for her decoding is the correct path at the end of the exclusive part.

\end{proof}

\subsubsection{Analysis for the One-Sided Setting}

\begin{proof}[Proof of \Cref{thm:listdecodingreduction13}]
As described in \Cref{sec:reductions} we assume that Alice wants to decode in the end and use the template from \Cref{sec:reductiontemplate} with $N'' = \frac{1.5}{\eps} N'$ rounds grouped into $b_1 = 1/\eps$ joint blocks and $b_2 = 1/\eps$ exclusive blocks in which Bob sends. We also choose $\eps_{ECC} < \eps$.

In order to show that Alice decodes to the correct path $\tau$ in the end we build on the proof of \Cref{thm:listdecodingreduction14}. In particular, we also show that the quantity $C' = c(\tau) - (C - c(\tau) - c(\emptyset))$ is positive for Alice in the end, which implies that $\tau$ is her majority path (with a confidence higher than all other non-empty paths combined). 

From the proof of \Cref{thm:listdecodingreduction14} we get that the contribution to $C'$ from the joint part is more than $\frac{1}{\eps} - \frac{4}{\eps} e_{ave,1} - 6$ where $e_{ave,1}$ is the average fraction of corruptions per block during the joint part. 

Since the joint part takes part during a $2/3$ fraction of the protocol the relative error rate in this part is at most $3/2 \cdot (1/3 - \eps) < 1/2 - \eps$ and the list decoder will have succeeded before the exclusive part of the protocol begins. The linearity of the confidence then shows again that the contribution of an exclusive block $i$ is at least $1 - 2 \eps - 2 e_i$. The contribution of the exclusive part towards $C'$ is therefore at least $\frac{1}{\eps} - 2 - \frac{2}{\eps}e_{ave,2}$ where $e_{ave,2}$ is the average fraction of corruptions per block during the exclusive part.

Together this gives 

\begin{align*}
C' &> \left(\frac{1}{\eps} - \frac{4}{\eps} \ e_{ave,1} - 6\right) + \left(\frac{1}{\eps} - \frac{2}{\eps}e_{ave,2} \ \ \ - 2\right)\\
   &= \ \ \,\frac{2}{\eps} - \frac{4}{\eps} \ \frac{E_1}{b_1 N'} \ \ \ \ \ \ \ \ \ \ \ \ \ \ \ - \frac{2}{\eps} \frac{E_2}{b_2 N'/2} \ - 8\\
   &= \ \ \,\frac{2}{\eps} - 4 \ \,\frac{E_1 + E_2}{N'} \ \ \ \ \ \ \ \ \ \ \ \ \ \ \ \ \ \ \ \ \ \ \ \ \ \ \ - 8.
\end{align*}

Here $E_1$ and $E_2$ stand for the total number of corruptions in the joint and exclusive part respectively. For an error rate of $1/3 - 2\eps$ the total number of errors satisfies:

$$E_1+E_2 \leq (1/3 - 2\eps) \frac{1.5}{\eps} N' = (\frac{1/2}{\eps} - 3)N'$$

which leads to 

\begin{align*}
C' &> \frac{2}{\eps} - 4 \left(\frac{1/2}{\eps} - \ \,3\right) \ - 8\\
   &= \frac{2}{\eps} - \ \ \ \ \ \ \frac{2}{\eps} \ \,+ 12 \ \ \ - 8 = 4 > 0.
\end{align*}

As desired, this shows that an error rate of at most $\frac{1}{3} - 2\eps$ results in Alice recovering the correct path $\tau$ by choosing the majority path. 
\end{proof}

\subsubsection{Analysis for the List Size Reduction of List Decoders}

\begin{proof}[Proof of \Cref{thm:listdecoding-listsizereduction}] 
recall that each party outputs $s'=O(1/\eps^2)$ many paths. Therefore, to complete the proof, it just remains to show that at least one of these paths is the correct path. Similar to the proof of \Cref{thm:listdecodingreduction14}, let $i^*$ be the first block in which the list decoder $\Pi'$ succeeds, that is, the first block after which the edge set $E_A \cup E_B$ contains the common path $\mathcal{P}$ of $\Pi$, and let $i'$ be the first block after block $i^*$ such that the error-rate in block $i'$ is at most $1/2-\eps$. Blocks $i^*$ and $i'$ exist as otherwise, every block except at most one must have error-rate greater than $1/2-\eps$. That would mean the total error-rate is at least $$(\frac{1}{\eps}-1) (\frac{1}{2}-\eps)/ (\frac{1}{\eps}) \geq (1-\eps)(\frac{1}{2}-\eps) >\frac{1}{2}-\frac{3\eps}{2} >\frac{1}{2}-2\eps,$$ which would be a contradiction. Now as the error-rate in block $i'$ is at most $1/2-\eps$ and since the edge-sets sent in this block already contain the common path (as $i'>i^*$), at least one the the decodings of each party is the correct path, which completes the proof.
\end{proof}

}

\section{Boosting List-Decoders}\label{sec:boosting}
In this section, we present a generic \emph{boosting} approach for improving the efficiency of list decoders, which can be viewed as a counterpart of \Cref{sec:boundedAdvBoosting}. In particular, the boosting that we present here improves the \emph{round complexity} (blow up) of the list decoders and it also allows us to generate \emph{computationally efficient} list-decoders, even from list-decoders with, e.g., exponential computational complexity. More concretely, as the basic boosting step, we explain that assuming a list-decoder coding scheme for $O(\log^2 n)$-round protocols, we can create a list-decoder coding scheme for $n$-round protocols with round complexity blow up similar to that of the $O(\log^2 n)$-rounds protocol and near-cubic computational complexity. A more advanced version with near-linear computational complexity appears in \Cref{sec:linearBoost}. We explain in \Cref{subsec:RecursiveBoost} how to recursively apply this boosting to get efficient list-decoders and then combine them with the reduction results to prove \Cref{thm:main14,thm:mainOthers}. For ease of readability we will use $\tilde{O}$-notation to hide $\log^{O(1)}n$ factors. 

\subsection{Basic Boosting Step: From Poly-logarithmic Protocols to Linear Protocols}\label{subsec:BasicBoost}
Here we show how to boost any list-decoder for protocols with $O(\log^2 n)$ rounds to a list-decoder for protocols with $n$ rounds, while loosing only an additive ${\eps'}$ term in the tolerable error rate and $\frac{1}{\eps'}$ factors in the round complexity and list size. More formally, we prove the following:

\begin{theorem}\label{thm:boosting} For any failure-exponent $C=\Omega(1)$, any $C'=\Omega(C)$, and any error-rate loss ${\eps'}$ such that $2\log{\frac{5}{{\eps'}}} \leq C\log^2 n$, the following holds: Suppose there is a list-decodable coding scheme that robustly simulates any $C'\log^2 n$-round protocol, while tolerating error rate $\rho$, and such that it has list size $s=\tilde{O}(1)$, round complexity $R C' \log^2 n$, computational complexity $T$, and failure probability at most $2^{-C\log ^2 n}$. Then, there exists a randomized list decoding coding scheme for $n$-round protocols that tolerates error rate $\rho-{\eps'}$ and has list size $s'=O(\frac{s}{{\eps'}})$, round complexity $O(\frac{R C'}{{\eps'}} \cdot n)$, computational complexity $\tilde{O}(\frac{n^3}{{\eps'}^{2}} \cdot T)$, and failure probability $2^{-Cn}$.
\end{theorem}

For simplicity, the reader might think of $C$ and $C'$ as large enough constants. Furthermore, the condition $2\log{\frac{5}{{\eps'}}} \leq C\log^2 n$ is a technical condition that is needed for the generality of this boosting but it is readily satisfied in all applications of interest in this paper.

\paragraph{Remark about the Computational Complexity of \Cref{thm:boosting}} For simplicity, in this section we present a boosting step that has computational complexity of $\tilde{O}(n^3)$. In \Cref{sec:linearBoost}, we present a more advanced version which has a computational complexity of $\tilde{O}(n)$. The modified boosting procedure from \Cref{sec:linearBoost} follows roughly the same outline as we present here except for modifying the working and computational complexity of a subprocedure used. 

\begin{proof}[Proof Outline of \Cref{thm:boosting}]
Let $\Pi$ be the original $n$-rounds protocol in the canonical form (see \Cref{sec:interactivecodingsettings}), let $\mathcal{T}$ be its binary tree in the canonical form, and let $E^{odd}$ and $E^{even}$ respectively represent the edges of $\mathcal{T}$ starting from odd and even levels. Furthermore, let $\mathcal{X} \subset E^{odd}$ and $Y\subset E^{even}$ respectively be the \emph{preferred edges} inputs of Alice and Bob. Finally, let $\mathcal{P}$ be the \emph{common path}\footnote{Refer to \Cref{sec:interactivecodingsettings} for the definitions of canonical form of protocols and the related concepts such as the format of the input or the common path.} in $\mathcal{X}\cup \mathcal{Y}$.  

The new coding scheme runs in $N= \frac{10 \, R C'}{{\eps'}} \cdot n$ rounds. These rounds are partitioned into $N'=\frac{10}{{\eps'}} \cdot \frac{n}{\log^2 n}$ \emph{meta-rounds}, each of length $R C' \log^2 n$ rounds. Furthermore, we break $\Pi$ into \emph{blocks} of length $\log^2 n$ rounds. In the simulation $\Pi'$ of $\Pi$, Alice and Bob always maintain edge-sets $\bar{E}_A$ and $\bar{E}_B$, respectively, which are rooted sub-trees of $\mathcal{T}$ and such that we have $\bar{E}_A \cap E^{odd} \subseteq \mathcal{X}$ and $\bar{E}_B \cap E^{even} \subseteq \mathcal{Y}$. Hence, always $\bar{E}_A \cap \bar{E}_B$ is a rooted sub-path of the common path $\mathcal{P}$. Initially, we have $\bar{E}_A=\bar{E}_B=\emptyset$. In the course of the simulation, we grow the edge-sets $\bar{E}_A$ and $\bar{E}_B$ by adding at most $s$ many blocks per meta-round. If a block added to $\bar{E}_A$ ends at a leaf of $\mathcal{T}$, then Alice adds a vote to this leaf, and Bob does similarly with respect to $\bar{E}_B$. We show that, at the end, if the total error-rate is less than $\rho-{\eps'}$, then for both Alice and Bob, the leaf of the common path is among the $s'=O(\frac{s}{{\eps'}})$ many leaves with the most votes.
 
Ideally, we would like each meta-round to simulate one block and if the error-rate is at most $\rho$, then this meta-round should make a progress of length $\log^2 n$ along the common path $\mathcal{P}$. That is, we would like that in each meta-round in which error-rate is at most $\rho$, $|(\bar{E}_A \cap \bar{E}_B) \cap P|$ grows by one block. However, realizing this ideal case faces one important challenge: in each meta-round, we cannot be sure of the correctness of the past blocks as the adversary could have corrupted them by investing enough errors. To remedy this issue, in each meta-round, the two parties first try to find the deepest block that has been computed correctly in the past; we call this the \emph{search phase}. Then the two parties simulate the next block extending from there downwards on $\mathcal{P}$; we call this the \emph{path-extension phase}. The search phase takes $\Theta(C\log^2 n)$ rounds while the path-extension phase takes $\log^2 n$ rounds. We choose the constants such that the total number of communications in search phase plus that of the path-extension phase is at most $C' \log^2 n$ rounds. This is doable because of the condition $C'=\Omega(10 C)$ in the statement of the lemma. Then, these $C' \log^2 n$ rounds of communication are wrapped in (and thus protected via) the list decodable coding scheme of $C' \log^2 n$ rounds, in the $R C' \log^2 n$ rounds of the meta-round. What we do on top of this list-decoder in each meta-round is as follows: for each meta-round, there are at most $s$ suggested transcripts. The parties add the extension blocks of these $s$ transcripts to their edge-sets $\bar{E}_A$ and $\bar{E}_B$ (but of course only if the block is consistent with the party's own local input $\mathcal{X}$ or $\mathcal{Y}$, otherwise the block gets discarded). Furthermore, for each of the $s$ transcripts, there is one path which is found in the search phase. If this path ends at a leaf, we add one \emph{vote} to this leaf. At the end of the whole simulation, each party outputs the $s'=O(\frac{s}{{\eps'}})$ leaves with the most votes. A pseudocode is presented in Algorithm \ref{alg:Booster}.

\begin{algorithm}[t]
\caption{Boosting List-Decoder, at Alice's Side}
\begin{algorithmic}[1]
\shortOnly{\small}
\State $\mathcal{X} \gets$ the set of Alice's preferred edges; 
\State $\bar{E}_A \gets \emptyset$; 
\State $N'=\frac{10}{{\eps'}} \cdot \frac{n}{\log^2 n}$; 
\For{$i=1$ to $N'$}

	\State Simulate the following $C' \log^2 n $-rounds protocol in $R\cdot C' \log^2 n$ rounds:
	\State \ \ \ \ \ \textbf{Search Phase}: \texttt{Find the deepest common path in $\bar{E}_A \cap \bar{E}_B$, let it be $P'$.}
	\State \ \ \ \ \ \textbf{Path-Extension Phase}: \texttt{Execute the protocol on the block extending $P'$.}
	\State $S \gets$ $s$ list-decodings of possible original outcomes of this protocol 
	\For {each outcome $\sigma \in S$} 
		\State $B \gets$ block executed in the path-extension phase of $\sigma$
		\If{$B$ is a path in $\mathcal{T}$ and $B \cap E^{odd} \subset \mathcal{X}$} 
			\State $\bar{E}_A \gets \bar{E}_A \cup \{B\}$
			\If {$B$ ends at a leaf $v$}
				\State $v.vote \gets v.vote+1$
			\EndIf
		\EndIf
	\EndFor

\EndFor
\State Output the $O(s/{\eps'})$ leaves with the most votes
\end{algorithmic}
\label{alg:Booster}
\end{algorithm}

In the above sketch, we did not explain how to solve the search phase. We abstract this phase as a new two party communication complexity problem over a noiseless channel, which we call the \emph{tree-intersection problem}, and discuss in \Cref{subsubsec:Tree-Intersection}. In particular, we explain how this problem can be solved in $O(\log^2 n)$ rounds with failure probability $1-2^{C\log^2 n}$, and with computational complexity of $\tilde{O}(n)$. Having solved the search phase, the complete the proof with a few simple arguments in \Cref{subsubsec:boostWrapUp}. In particular, we show that in each meta-round in which the error-rate is less than $\rho$, with probability at least $1-2^{C\log^2 n}$, either $|(\bar{E}_A \cap \bar{E}_B) \cap P|$ grows by at least one block or if $|(\bar{E}_A \cap \bar{E}_B) \cap P|$ already contains a leaf, then this leaf receives one more vote. We then show that with probability at least $1-2^{-Cn}$, the leaf at the end of the common path $\mathcal{P}$ receives at least $\Theta(N' {\eps'})$ votes. On the other hand, each of $\bar{E}_{A}$ and $\bar{E}_B$ can contain a total vote of at most $N'\cdot s$. Hence, we get that the correct path is among the $s'=O(\frac{s}{{\eps'}})$ leaves with the most votes, with probability at least $1-2^{-Cn}$.
\end{proof}

\subsubsection{The Tree-Intersection Problem} \label{subsubsec:Tree-Intersection}
\begin{definition}(\textbf{The Tree-Intersection Problem}) Suppose that Alice and Bob respectively have edge sets $\bar{E}_A$ and $\bar{E}_B$ that correspond to subtrees of a complete binary tree $\mathcal{T}$ of depth $n$ rooted at the root of $\mathcal{T}$, and that $|\bar{E}_A|\leq M$ and $|\bar{E}_B|\leq M$, where $M = \tilde{O}(n)$. Now, given the promise that $P = \bar{E}_A \cap \bar{E}_B$ is a path, Alice and Bob want to recover the path $P$ with as little communications over a noiseless binary channel as possible, while failing to so only with negligible probability.
\end{definition}

\begin{figure}[t]
	\centering
		\fullOnly{\includegraphics[width=0.8\textwidth]{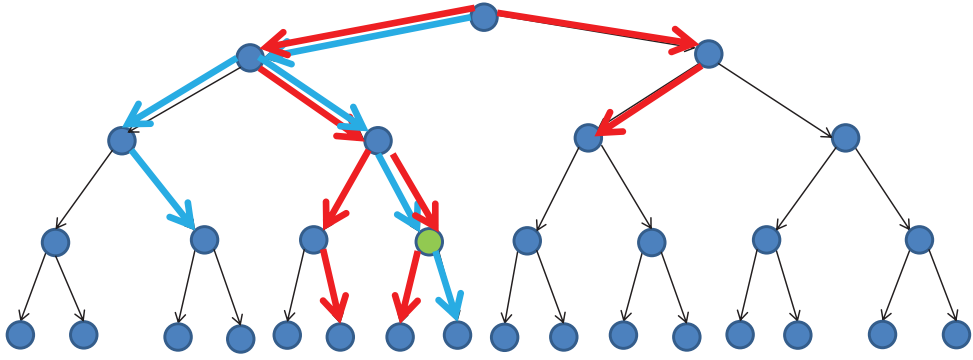}}
		\shortOnly{\includegraphics[width=0.6\textwidth]{Tree-Intersection}}
	\caption{The Tree-Intersection Problem}
	\label{fig:tree-intersection}
\end{figure}

\Cref{fig:tree-intersection} shows an example of this problem where edges in $\bar{E}_A$ and $\bar{E}_B$ are indicated with blue and red arrows, respectively, and the path $P$ is the path from the root to the green node.

We present a simpler $O(C\log^2 n)$ rounds solution, in \Cref{lem:tree-intersection}, which explains the main approach but has failure probability at most $2^{-\Omega(C\log n)}$. In \Cref{lem:tree-intersection2}, we explain a better version with similar round complexity but failure probability at most $2^{-\Omega(C\log^2 n)}$. 
 
\begin{lemma}\label{lem:tree-intersection}For any $C$, there is a tree-intersection protocol which uses $O(C \log^2 n)$ rounds of communication on a binary channel, $O(C\log n)$ bits of randomness, and polynomial-time computation, and finds path $P$ with failure probability at most  $2^{-C\log n}$. This protocol has computation complexity $\tilde{O}(C M)$.
\end{lemma}
\begin{proof}
Alice samples $\Theta(C\log n)$ random bits and sends them to Bob. This defines a hash function $h:\{0,1\}^{n}\rightarrow \{0,1\}^{\Theta(C\log n)}$. Choosing large enough constants, we have that the probability that there are two different paths (among the paths of Alice and Bob) that have equal hash-values is at most $M^2 \cdot 2^{-\Theta(C\log n)} \leq 2^{-C\log n}$, where the $M^2$ is for a union bound over all pairs. In the rest, we assume that no two different paths have equal hash values and we use this hash in a binary search for finding the intersection $P = \bar{E}_A \cap \bar{E}_B$. 

Alice finds an edge $e\in \bar{E}_A$ that cuts her edge-set $\bar{E}_A$ in two ``semi-balanced" parts, each containing at least $|\bar{E}_A|/3$ edges. That is, an edge $e$ such that the following holds: let $\mathcal{T}_e$ be the subtree of $\mathcal{T}$ below edge $e$. Then, edge $e$ should satisfy $|\bar{E}_A|/3 \leq |\bar{E}_A \cap \mathcal{T}_e| \leq 2|\bar{E}_A|/3$. Note that such an edge $e$ exists and also can be found in $\tilde{O}(C M)$ time. Once Alice finds such an edge $e$, she then sends $h(P_e)$ to Bob, where $h(P_e)$ is the hash-value of the path $P_e$ starting from the root and ending with edge $e$. Bob checks whether he has a path with the same hash-value $h(P_e)$ and reports the outcome to Alice by sending one bit. If there is a path with matching hash value, then $e$ is construed to belong to the common path. Otherwise, if there is no such path with matching hash-value, this is construed as $e$ not belonging to the common path. In either case, Alice can discard at least a $1/3$ fraction of $\bar{E}_A$. This is because, if $e$ is not on the common path, then every edge in $\bar{E}_A \cap \mathcal{T}_e$ can be discarded. On the other hand, if $e$ is on the common path, then we are sure that the path starting from the root and ending with $e$ is in the common path. Thus, edges on this path can be also ignored from now on as certainly being on the path and the remaining problem is to only solve the tree-intersection in $T_e$. Note that any edge that diverges from $P_e$ before $e$ gets discarded as well as it cannot be on the common path $P$.

Iterating the above step $\log_{3/2} n$ times leads to Alice finding the common path. Alice can then report this path to Bob by just sending the related hash value. The whole procedure succeeds if the hash-values of different paths in $\bar{E}_A\cup \bar{E}_B$ are different which as discussed before happens with probability at least $1-2^{-C \log n}$.
\end{proof}

To reduce the failure probability to $2^{-\Omega(\log^2 n)}$, the key change is that we use a probabilistic binary search approach instead of the deterministic binary search used above. The main point is to try to cover for the possibility that each hash-value checking step can fail with probability $2^{-\Theta(\log n)}$ by allowing backtracking in the binary search. We note that getting this better $2^{-\Omega(\log^2 n)}$ failure probability, that is a failure probability that is exponential in the communication complexity of the tree-intersection solution, is most interesting for our non-uniform deterministic coding schemes. For the randomized ones, even if we just use the tree-intersection explained above that has failure probability $2^{-\Omega(\log n)}$, we get a final list-decoder that has failure probability $2^{-\Omega(n/\log n)}$ which is still quite small. To simplify the exposition, we defer the details of the improved tree-intersection (that has 
failure probability $2^{-\Omega(\log^2 n)}$) to \Cref{subsub:deterministicBoost}, where we discuss our non-uniform deterministic coding schemes.

\subsubsection{Completing the Basic Boosting Step}\label{subsubsec:boostWrapUp} 
We now complete the proof of \Cref{thm:boosting}. We first show that for each meta-round with small error-rate, this meta-round either makes a block of progress on the common path or it adds a vote to the leaf at the end of the common path.
\begin{lemma}\label{lem:metaround} In each meta-round in which error-rate is at most $\rho$, with probability at least $1-O(2^{-C\log^2 n})$, either $|\bar{E}_A\cap \bar{E}_B \cap \mathcal{P}|$ increases by $\log^2 n$ or one vote is added to the leaf at the end of $\mathcal{P}$, on both of Alice and Bob's sides.
\end{lemma}
\begin{proof}
Note that in the absence of errors, each meta-round would with probability at least $1-2^{-2C\log^2 n}$ find the deepest path in $\bar{E}_A \cap \bar{E}_B$ and then extend it by one block along $\mathcal{P}$ (if it already does not end in a leaf). The list-decoding coding scheme for $C' \log^2 n$ round protocols provides the following guarantee: if the error-rate in this meta-round is at most $\rho$, with probability at least $1-2^{-2C\log^2 n}$, we get a list of $s$ possible transcripts of this $\log^2 n$ round protocol, one of which is correct. In the algorithm, we add all of the $s$ possible new blocks, one for each transcript, to $\bar{E}_A$ and $\bar{E}_B$, and also if the blocks end at a leaf, we add one vote to the respective leaf. Hence, if the meta-round has error-rate at most $\rho$, with probability at least $1-O(2^{-2C\log^2 n})$, either $|\bar{E}_A\cap \bar{E}_B \cap \mathcal{P}|$ increases by one block or each of Alice and Bob add one vote to the leaf at the end of $\mathcal{P}$.
\end{proof}

\begin{proof}[Proof of \Cref{thm:boosting}]
Let us call a meta-round bad if one of the following holds: (a) its error-rate is greater than $\rho$, (b) its error-rate is less than $\rho$ but the parties neither make one block of progress along $\mathcal{P}$ together nor they both add a vote to the leaf at the end of $\mathcal{P}$. At most $\frac{\rho-{\eps'}}{\rho}$ fraction of the meta-rounds have error-rate greater than $\rho$. On the other hand, \Cref{lem:metaround} tell us that in each meta-round with error-rate at most $\rho$, with probability at least $1-O(2^{-2C\log^2 n})$, parties either both make one block of progress along $\mathcal{P}$ or both add a vote to the leaf at the end of $\mathcal{P}$. Thus, with probability at least $1-2^{-Cn}$, the number of bad meta-rounds in which error-rate is less than $\rho$ is at most ${\eps'} N'/2$. This is because, the probability that there are more such meta-rounds is at most 
%
$$\sum_{i={\eps'} N'/2}^{N'} \binom{N'}{i} O(2^{-C \, \log^2 n})^i \leq \sum_{i={\eps'} N'/2}^{N'} (\frac{5}{{\eps'}})^i \cdot 2^{-i C \, \log^2 n)} \leq \sum_{i={\eps'} N'/2}^{N'} 2^{i (\log{\frac{5}{{\eps'}}} -C  \log^2 n)} \leq 2^{-\frac{N'{\eps'}}{4} \cdot C \cdot \log^2 n},$$
%
which is less than or equal to $2^{-Cn}$. Hence, with probability at least  $1-2^{-Cn}$, the fraction of bad meta-rounds is at most $\frac{\rho-{\eps'}}{\rho} +\frac{{\eps'}}{2} \leq 1-{\eps'}/2$. Therefore, there are at least $N' \cdot \frac{{\eps'}}{2}$ good meta-rounds. Note that each good meta-round either extends the common path along $\mathcal{P}$ by one block or adds a vote to the leaf at the end of $\mathcal{P}$. On the other hand, at most $\frac{n}{\log^2 n} \leq \frac{N'{\eps'}}{4}$ meta-rounds can be spent on extending the common path along $\mathcal{P}$ by one block each. Hence, the leaf at the end of $\mathcal{P}$ receives at least $\frac{N'{\eps'}}{2} -\frac{N'{\eps'}}{4} \geq \frac{N'{\eps'}}{4}$ votes. On the other hand, each of $\bar{E}_{A}$ and $\bar{E}_B$ can contain at most $N'\cdot s$ votes. Therefore, with probability at least $1-2^{-Cn}$, the correct path is among the $O(\frac{s}{{\eps'}})$ leaves with the most votes.
\end{proof}

\fullOnly{\subsection{Recursive Boosting}\label{subsec:RecursiveBoost}
Here, we explain that recursively applying the boosting step provides us with efficient list-decoders, even starting from non-efficient ones. In \Cref{lem:recBoosting}, we present the result that allows us to obtain list-decoders with linear communication complexity and near-linear computational complexity using the  linear communicationl complexity but exponential computational complexity coding schemes of Braverman and Rao\cite{BR11} and Braverman and Efremenko\cite{BE14}. Note that \cite{BR11} provides a unique decoder tolerating error rate $1/4-\eps$ and \cite{BE14} provides a list decoder tolerating error-rate $1/2-\eps$. 
As an alternative path, in \Cref{thm:logstar-boosted-simplified}, we show how to obtain a list-deocder for error-rate $1/2-\eps$ with almost linear communication complexity of $N=n \cdot 2^{O(\log^* n \, \cdot \, \log{\log^* n})}$ rounds and with near-linear computational complexity just by recursively boosting the simple quadratic communication complexity list decoder of \cite{GHS13}. 

\begin{lemma}\label{lem:recBoosting} Suppose that there is a list-decodable coding scheme that robustly simulates any $\Theta((\log \log \log n)^2)$-round protocol, tolerates error-rate $\rho$, and has a constant size alphabet, round complexity $O((\log \log \log n)^2)$, failure probability $o(1)$, a constant list size and computational complexity of $\tilde{O}(1)$. Then, for any $\eps>0$, there is a list-decodable coding scheme that robustly simulates any $n$-round protocol tolerating error rate $\rho -\eps$, with a constant size alphabet, round complexity $O(n)$, failure probability $2^{-\Omega(n)}$, a constant list size of $O(1/\eps^2)$ and computational complexity of $\tilde{O}(n)$.
\end{lemma}
\begin{proof}
We apply \Cref{thm:boosting} to the list decodable coding scheme at hand for three times, with constant $\eps'=\eps/6$, and large enough constants $C$ and $C'$. The first boosting step takes us to coding schemes for protocols of length $O((\log \log n)^2)$. Note that this is because $\log(O(\log^2\log n)) = O(\log \log \log n)$. Then, the second boosting takes us to coding schemes for protocols of length $O(\log n)^2$, and finally the third one takes us to a list-decodable coding scheme for $n$-round protocols. For each boosting step, we sacrifice a factor of $O(\frac{1}{\eps} \log(\frac{1}{\eps}))$ factor in the round-complexity: a $O(\frac{1}{\eps})$ comes directly from \Cref{thm:boosting} and then transforming the result into canonical form gives the $O(\log(1/\eps))$ factor. Furthermore, in each boosting step, the list size grows by $O(1/\eps)$. Also, in each recursion level, the tolerable error-rate decreases by $\frac{\eps}{6}$ which after $3$ recursions, takes us to tolerable error-rate of $\rho-\eps/2$. Hence, at the end, we get a list-decoder for $n$-round protocols with round complexity $O(\frac{n}{\eps^3 \log^3(1/\eps)})=O(n)$, constant list size of $O(1/\eps^3)$, failure probability $2^{-\Theta(n)}$, and computational complexity $\tilde{O}(n)$. At the end, we apply \Cref{thm:listdecoding-listsizereduction} which reduces the list size to $O(\frac{1}{\eps^2})$. 
\end{proof}

\begin{lemma}\label{thm:logstar-boosted-simplified}
For any constant $\eps>0$, there is a randomized list-decodable coding scheme that robustly simulates any $n$-round protocol over any channel with constant alphabet size $O(1/\eps)$ and 
error rate at most $1/2 - \eps$, in $n \cdot 2^{O(\log^* n \, \cdot \, \log{\log^* n})}$ rounds,
with list size $O(1/\eps^2)$, computational complexity $\tilde{O}(n)$, and failure probability at most $2^{-\Omega(n)}$. 
\end{lemma}
\begin{proof} [Proof of \Cref{thm:logstar-boosted-simplified}]
Let $C=\Theta(\log^* n)$ and $C'=\Theta(C)$ for large enough constants that satisfy the conditions of \Cref{thm:boosting}. From \cite[Theorem 3.4]{GHS13}, we get a deterministic list decoder that robustly simulates any $O(\log^*{n})^2$-rounds protocol over a channel with alphabet size $O(1/\eps)$ and error rate $1/2-\eps/2$, such that it has list size $O((1/\eps)^2)$, round complexity $O((\log^{*}{n})^4)$, and computational complexity $O((\log^{*}{n})^4)$. Then, we recursively apply \Cref{thm:boosting} for at most $k = \floor{\log^* n}$ times, with parameter ${\eps'} = \frac{\eps}{2k}$. In particular, the first recursion takes us to coding schemes for protocols of length $2^{\Theta(\log^*)}$, the next recursion takes us to coding schemes for protocols of length $2^{\Theta(2^{\log^* n})}$ and so on. After at most $\ceil{\log^* n}$ recursions, we get a coding scheme for $n$-rounds protocols. For each recursion level, we sacrifice a factor of $O(C'/{\eps'}) = O(\frac{(\log^*n)^2}{\eps})$ factor in the round-complexity, and a factor of $O(1/{\eps'})=O(\frac{\log^* n}{\eps})$ in the list size. Also, in each recursion level, the tolerable error-rate decreases by $\frac{\eps}{2k}$ which after $k$ recursions, takes us to tolerable error-rate of $1/2-\eps$. We note that \Cref{thm:boosting} assumes that the protocol to be used for each block is in canonical form but the protocol that it generates does not have this form. However, as noted in \Cref{sec:interactivecodingsettings} one can easily transform any protocol $\Pi$ into one of  canonical form while increasing its round complexity by at most $O(\log{\sigma(\Pi)})$, where $\sigma(\Pi)$ is the alphabet size of $\Pi$. This means the round complexity increases by $o(\frac{1}{\eps})$ per recursion level. At the end of $k$ recursions, we get a randomized list decoder that robustly simulates any $n$-rounds protocol over a channel with alphabet size $O(1/\eps)$ and error rate $1/2-\eps$, such that it has round complexity $n \cdot (\frac{\log^* n}{\eps})^{O(\log^* n)}$ , list size $(\frac{\log^* n}{\eps})^{O(\log^* n)}$, computational complexity $\tilde{O}(n \cdot \frac{1}{\eps^{\log^* n}})$ and failure probability at most $2^{-Cn} = 2^{-\omega(n)}$. For the regime of our interest where $\eps$ is a constant, these bounds can be simplified as follows: round complexity $n \cdot (\log^* n)^{O(\log^* n)}$, list size $(\log^* n)^{O(\log^* n)}$, computational complexity $\tilde{O}(n)$ and failure probability $2^{-\omega(n)}$. At the end, we apply \Cref{thm:listdecoding-listsizereduction} which reduces the list size to $O(\frac{1}{\eps^2})$. 
\end{proof}

\subsection{Deterministic Boosting}\label{subsub:deterministicBoost}
In \Cref{subsubsec:Tree-Intersection}, we explained a tree-intersection algorithm with round complexity of $O(C\log^2 n)$ and failure probability at most $2^{-\Omega(C\log n)}$. Here, we first explain how to reduce the failure probability to $2^{-\Omega(\log^2 n)}$. The key change is that we now use a probabilistic binary search approach, which tries to recover for the possibility that each hash-value checking step can fail with probability $2^{-\Theta(\log n)}$. 

We first present the probabilistic binary search approach in a more general form in \Cref{lem:probabilisticBS}. Then, in \Cref{lem:tree-intersection2}, we explain how applying this probabilistic binary search reduces the failure probability of the tree-intersection protocol to $2^{-\Omega(\log^2 n)}$. 

\begin{lemma}\label{lem:probabilisticBS} Consider a binary search tree with depth $h$, where we want to find the leaf with value $x$. Suppose that each comparison with each vertex $v$ in this binary search gives an incorrect output with probability at most $\delta < 0.01$, with independence between failures of different comparisons. Then, for any $C \geq 1$, there is a probabilistic binary search protocol that runs in $O(C h)$ steps and finds the correct leaf with probability at least $1 - 2^{-C\log\frac{1}{\delta} h}$. 
\end{lemma}

\begin{proof}
The probabilistic binary search works as follows: We first modify the binary search tree by hanging a chain of length $10 C h$ from each leaf. Then, consider a step of the binary search, and suppose we are now at a node $v$ of the tree, which means that we believe $x$ to be in the subtree rooted in node $v$. We first ``double-check" that $x$ is indeed in the sub-tree rooted at $v$. This can be done for example by two comparisons with respect to the smallest and largest values in the subtree rooted in $v$. If the double-check fails, we backtrack to parent of $v$ in the binary search tree. On the other hand, if the check passes, we do as follows: if $v$ has only one child---that is, if $v$ is node in the chain hanging from a leaf---, then we simply move one step downwards on the chain. On the other hand, if $v$ has at least two children, then we compare $x$ with the value of $v$ and move to the left or right child of $v$ accordingly. At the end of $10 C h$ steps, if we are in the chain hanging from a leaf $u$, we output $u$. Otherwise, we output an arbitrary answer.

For the analysis, we direct all the edges towards the deepest node in the chain hanging from the leaf that has value $x$. A move along an edge is called \emph{correct} if it accords with the direction of the edge and it is called \emph{incorrect} otherwise. It is easy to see that if the number of correct moves minus that of the incorrect moves is greater than $h$, then the output is correct. In other words, if the number of incorrect moves is less than $5Ch - \frac{h}{2}$, then the output is correct. Using a union bound, we see that the probability of an incorrect move is at most $3\delta$ as we use at most $3$ comparisons in each step, two for the double-check and one for comparing versus the node itself. Hence, the probability that the output is incorrect is at most 
\begin{eqnarray}
\sum_{i=5 Ch - \frac{h}{2}}^{10 Ch} \binom{10 Ch}{i} (3\delta)^{i} \leq \sum_{i=5 Ch - \frac{h}{2}}^{10 Ch} 2^{3 i} \cdot 2^{-i\log\frac{1}{3\delta}} \leq 2^{-(5 Ch-\frac{h}{2}) \cdot \frac{\log\frac{1}{\delta}}{4}} \leq 2^{-C\log\frac{1}{\delta} h}. \nonumber
\end{eqnarray}
\end{proof}

\begin{lemma}\label{lem:tree-intersection2} For any $C=\Omega(1)$, there is a tree-intersection protocol which uses $O(C\log^2 n)$ rounds of communication over a noiseless binary channel and $O(C\log^2 n)$ bits of randomness, and finds path $P$ with failure probability at most  $2^{-C \log^2 n}$. This protocol has computation complexity $\tilde{O}(M)$.
\end{lemma}
\begin{proof} Proof follows essentially by combining \Cref{subsubsec:Tree-Intersection} with \Cref{lem:probabilisticBS}. That is, instead of the deterministic binary search applied in \Cref{lem:tree-intersection}, we use a probabilistic binary search for $\Theta(C\log n)$ steps. Again, Alice is the main party that performs the binary search (which is now probabilistic), while Bob only checks hash values and responds whether he has a match or not. This is a binary search tree with depth $h=\log_{3/2} M =\Theta(\log n)$. Furthermore, we use a new hash function for each hash test, where Alice sends the $O(\alpha' \log n)$ bits of the seed of the hash function and the value of hash evaluated on a path to Bob and then Bob responds with a one bit answer indicating whether he has a match or not. Here $\alpha'$ is a large enough constant. Thus, the probability of a mistake in each step of binary search is $\delta = 2^{-\alpha'\log n}$.

The probabilistic binary search applied to the tree-intersection works as follows: Consider a step of the binary search, and suppose we are now at a node $v$ of the tree. We first ``double-check" the path ending at $v$ using hashing. That is, Alice sends the hashing seed $r_i$ and the hash function $h^i(P_v)$ to Bob and receives back whether Bob has any match or not. If the check fails and there is no match, Alice backtracks to the last node that she checked before $v$, that is, the ``parent" of $v$ in the binary search tree. On the other hand, if the check passes and Bob has a match for $h(P_v)$, then Alice moves to testing the next node $w$ in the sub-tree under $v$ that gives an \emph{semi-balanced} partition of the $\bar{E}_A$ edges below $v$. That is, Alice sends the seed $r_{i'}$ and the hash value of $h^{i'}(P_{w})$ and receives from Bob whether he has a path with matching hash-value or not. Depending on whether there is a match or not, Alice moves to examining the next node. Every time that Alice is in a node that does not have any downwards $\bar{E}_A$ edge going out of it---that is, a leaf of the binary search tree---with every check of the corresponding path, she either increases or decreases the counter of this leaf by one, depending on whether the check succeeded or failed. If the counter is equal to zero and test fails, Alice backtracks to the parent of this node in the binary search. This counter corresponds to walking on the chain hanging from this leaf of the binary search tree, as explained in \Cref{lem:probabilisticBS}. At the end, the output is determined by the node in which Alice resides after $\Theta(C \log n)$ steps of the search. 

The analysis is as in \Cref{lem:probabilisticBS}: Putting $\delta = 2^{-\alpha'\log n}$ and $h=\Theta(\log n)$, we get that the binary search succeeds with probability at least $1- 2^{-C \log^2 n}$.
\end{proof}

This small failure probability of $1- 2^{-C \log^2 n}$ allowed us to prove \Cref{thm:boosting}. Now using the strong property that \Cref{thm:boosting} provides exponentially small failure probability with arbitrarily large failure exponent (at the cost of larger constant factor in the round complexity), we can also show  that coding scheme continues to work if all randomness is known to the adversary (in advance). This also implies computationally efficient nonuniform deterministic list-decoders with the same tolerable error rate and communication complexity.

\begin{theorem}\label{lem:deterministic-boosting} For any error-rate loss ${\eps'}$, the following holds: Suppose there is a deterministic list-decodable coding scheme that robustly simulates any $C'\log^2 n$-rounds protocol, while tolerating error rate $\rho$, and such that it has list size $s=\tilde{O}(1)$, and round complexity $R C' \log^2 n$, where $C'=\Omega(\frac{s}{{\eps'}})$ and $\log{\frac{1}{{\eps'}}} =O(C'\log^2 n)$. Then, there exists a deterministic list decoding coding scheme for $n$-round protocols that tolerates error rate $\rho-{\eps'}$ and has list size $s'=O(\frac{s}{{\eps'}})$ and round complexity $O(\frac{R C'}{{\eps'}} \cdot n)$.

\end{theorem}
\begin{proof}
We argue that the randomized coding scheme described in the proof of \Cref{thm:boosting} has a positive chance of being successful against all behaviors of adversary. This shows that there exists a way to fix the randomization used by this coding scheme and make it deterministic such that it succeeds against all behaviors of adversary. The proof is essentially by doing a union bound argument, but we can not afford to do this union bound over all behaviors of the adversary. Instead, we count the number of possibilities of the ``effect" of the behavior of the adversary and do the union bound over this count. More concretely, this counting is as follows: recall that each meta-rounds of the algorithm used in \Cref{thm:boosting} uses a list-decoder coding scheme, which generates a list of $s$ possible transcripts. Regarding each of these transcripts, what we care about is only the block found in the path-extension phase. For this block, there are at most $n^{O(1)} 2^{\log^2 n} \leq 2^{2\log^2 n}$ valid possibilities. This is because, there are at most $n^{O(1)}$ ways to pick the starting node of the block as it has to be within one of the nodes that is an endpoint of an edge in $\bar{E}_A$ for Alice or $\bar{E}_B$ for Bob and then there are at most $2^{\log^2 n}$ ways to extend it by a path of length $\log^2 n$, i.e., a potential block. Hence, there are at most $2^{2s\log^2 n}$ different combinations for the $s$ blocks that are provided by the list-decoder coding scheme. This means for the $N'$ meta-rounds of the simulation, there are at most $2^{2N's \log^2 n}= 2^{\frac{20s}{{\eps'}} \cdot}$ many possibilities for the blocks provided by the list-decoders of the meta-rounds, and this captures all the effect of the behavior of the adversary. If we choose $C$ of \Cref{thm:boosting} large enough, e.g., $C=\frac{30s}{{\eps'}}$, we get that the probability that there exists at least one behavior of the adversary which makes the randomized algorithm fail is at most $2^{2N's \log^2 n} 2^{-Cn}= 2^{(\frac{20s}{{\eps'}}- \frac{30s}{{\eps'}}) \cdot n} \leq 2^{-Cn/3} \ll 1$. That is, there exists a fixing of the randomization used in the algorithm that makes it robust against any adversary.
\end{proof}

As a result, we can obtain non-uniform deterministic variants of \Cref{lem:recBoosting,thm:logstar-boosted-simplified} by recursively applying the deterministic boosting of \Cref{lem:deterministic-boosting} instead of the randomized boosting of \Cref{thm:boosting}.

\begin{corollary}\label{crl:recBoosting-deterministic}
Suppose that there is a deterministic list-decodable coding scheme that robustly simulates any $\Theta((\log \log \log n)^2)$-round protocol, tolerates error-rate $\rho$, and has a constant size alphabet, round complexity $O((\log \log \log n)^2)$, a constant list size and computational complexity of $\tilde{O}(1)$. Then, for any $\eps>0$, there is a non-uniform deterministic list-decodable coding scheme that robustly simulates any $n$-round protocol, tolerates error rate $\rho -\eps$, and has a constant size alphabet, round complexity $O(n)$, a constant list size of $O(1/\eps^2)$ and computational complexity of $\tilde{O}(n)$.
\end{corollary}
\begin{corollary}\label{crl:logstar-boosting-deterministic}
For any constant $\eps>0$, there is a non-uniform deterministic list-decodable coding scheme that robustly simulates any $n$-round protocol over any channel with constant alphabet size $O(1/\eps)$ and 
error rate at most $1/2 - \eps$, in $n \cdot 2^{O(\log^* n \, \cdot \, \log{\log^* n})}$ rounds,
with list size $O(1/\eps^2)$, and computational complexity $\tilde{O}(n)$.
\end{corollary}

}
\fullOnly{\section{Boosting with Near-Linear Computational Complexity}\label{sec:linearBoost}

In this section, we improve the boosting approach presented in \Cref{sec:boosting} to produce list-decoders with near linear instead of cubic computational complexity. We advise the reader to read \Cref{subsec:BasicBoost} before this section. We achieve this speedup by designing an efficient data structure ontop of which the tree-intersection problem can be solved in $\tilde{O}(1)$ instead of $\tilde{O}(n^2)$ computation steps. 
The exact theorem we prove in this section is a direct equivalent to \Cref{lem:tree-intersection2} except with a significantly improved computational complexity:

\begin{theorem}
There is an incremental data structure that maintains a rooted subtree of the rooted infinite binary tree under edge additions with amortized computational complexity of $\tilde{O}(1)$ time per edge addition. Furthermore, for any $C = \Omega(1)$ and given two trees maintained by such a data structure, there is a tree-intersection protocol that uses $O(C\log^4 n)$ rounds of communication over a noiseless binary channel, $O(C\log^4 n)$ bits of randomness, and $\tilde{O}(1)$ computation steps to solve the tree-intersection problem, that is, find the intersection path, with failure probability at most $2^{-C \log^4 n}$.
\end{theorem}



We first provide an overview over the computational bottlenecks in our simple tree-intersection algorithm from \Cref{sec:nearlinearchallenges} which identifies two challenges to overcome. We in \Cref{sec:outlinedoublebinarysearch,sec:outlinefasthashing} we give the ideas, algorithms, and proofs addressing each challenge respectively. 

\subsection{Challenges}\label{sec:nearlinearchallenges}

The boosting algorithm in \Cref{subsec:BasicBoost} consists of $\tilde{O}(n)$ iterations in which an tree-intersection problem is solved using $\tilde{O}(n)$ hash computations and comparisons which each take $\tilde{O}(n)$ time to compute. To bring this $\tilde{O}(n^3)$ running time down to $\tilde{O}(n)$ we show how to utilize efficient pre-computations to solve each tree-intersection problem with $\tilde{O}(1)$ hash computations each running in $\tilde{O}(1)$ time. 

The reason for the $\tilde{O}(n)$ hash computations and comparisons comes from the following part of the simple tree-intersection algorithm from \Cref{sec:boosting}: Each tree-intersection consists of Alice performing a $O(\log n)$ step binary search on her edge-set $\bar{E}_A$, each time asking whether a given path $P_e \subseteq \bar{E}_A$ starting from the root and ending in edge $e$ is also present in Bob's edge-set $\bar{E}_B$. For this, Alice sends an $\tilde{\Theta}(1)$ long hashing of $P_e$ to Bob, relying on the fact that, with high probability, this hash will be unique among all paths both edge sets. However, in order for Bob to find out whether Alice's path $P_e$ is contained in $\bar{E}_B$ he needs to compute the hash for each of his paths $P'_e \subseteq \bar{E}_B$ and check whether the hashes matches. Since in most iterations $|\bar{E}_B| = \tilde{O}(n)$ this requires $\tilde{O}(n)$ hash computations and comparisons. The first change necessary is therefore to perform a different search procedure, which performs only $\tilde{O}(1)$ hash computations and comparisons, which we give in \Cref{sec:outlinedoublebinarysearch}.

The harder bottleneck to address is the computation of the hashes themselves. Since most paths to be checked for equality will be $\tilde{\Theta}(n)$ long computing a computing a hash requires $\tilde{\Omega}(n)$ rounds. Since even a single symbol difference between two strings is supposed to be found with at least constant probability it is actually far from clear that one can do better overall. In \Cref{sec:outlinefasthashing} we follow ideas from \cite{BN13} and show that with the right coding-precomputations, which can be done using an efficient data structure, each hash comparison can be performed in $\tilde{O}(1)$ time.

\subsection{Double Binary Search}\label{sec:outlinedoublebinarysearch}

The alternative search approach we take for reducing the number of hash comparisons per tree-intersection to to $\tilde{O}(1)$ is to do two binary searches, one on Alice's side and the other on Bob's side. More concretely, once Alice has fixed a rooted-path $P_e \subseteq \bar{E}_A$, which starts from the root and ends at an edge $e$, that she wants to check whether $P_e \subseteq \bar{E}_B$ or not, Bob performs a binary search to help. This binary search will be similar to the one that Alice uses: Each time Bob picks an edge $e'\in \bar{E}_B$ and Bob sends to Alice the depth $\ell$ of this edge and the hashing of the path $P_{e'}$ starting from the root and ending in $e'$. Now Alice simply compares the received hash value with the hash of the path $P^\ell_e$ which is the prefix of length $\ell$ of $P_e$. If the hash functions match, the parties know that (w.h.p.), $P^\ell_e \subseteq \bar{E}_B$. Then, Bob will proceed to search for a longer match in the sub-tree of $\bar{E}_B$ below $e'$. If the hash values do not match, the parties know that $P^\ell_e \not\subseteq \bar{E}_B$. Thus, Bob will (temporarily) discard the sub-tree of $\bar{E}_B$ below $e'$ and search for a match in the remainder of $\bar{E}_B$. Every time, Bob chooses the edge $e'$ such that in either case, at least a $1/3$ of the active-remaining part of $\bar{E}_B$ gets discarded. With this approach, in each tree-intersection problem, there will be only $O(\log^2 n)$ hash-value pairs that are compared with each other. A $O(\log n)$ factor comes from Alice's binary search for her various choices of $P_e \in \bar{E}_A$ and the other $O(\log n)$ factor comes from Bob's binary search for his choices of edge $e' \in \bar{E}_B$. This already brings the total number of hash-function comparisons over all meta-rounds to $\tilde{O}(n)$. We later explain that for each of these $O(\log^2 n)$ hash-value comparisons, we use $O(\log^2 n)$ rounds of communication, thus making the whole tree-intersection solution an $O(\log^4 n)$-round protocol, which fits in the meta-round and can be protected via the list-decoder for $O(\log^4 n)$-round protocols. We note that for these binary searches on each side, one can either use the simpler deterministic binary search, or the more advanced probabilistic binary search explained in \Cref{lem:tree-intersection2} to get the better failure probability of $2^{-\Omega(n)}$for the final result of boosting.

\subsubsection{The Data Structures}

to make this outline an algorithm with computational complexity of $\tilde{O}(n)$, there are three main elements that remain to be explained, which we describe next. The way to achieve these will be through constructing a number of \emph{data structures}, which we explain later.

\begin{itemize}
\item[(I)] The first element is that Alice needs to be able to find the search points $e\in \bar{E}_A$---or equivalently the rooted-path $P_e \subseteq \bar{E}_A$--- in $\tilde{O}(1)$ time. The same is true about the search points $P_{e'} \subseteq \bar{E}_A$ of Bob.

\item[(II)] The second element is that, once Bob gives a length $\ell$ to Alice, Alice needs to be able to find $P_e^{\ell}$---that is, the prefix of length $\ell$ of $P_e$---in amortized $\tilde{O}(1)$ time. In other words, given an edge $e$ and a length $\ell'=|P_e|-\ell$, Alice should be able to find the edge $e''$ that is $\ell'$ hops above $e$ on the path connecting $e$ to the root, i.e., the last edge on $P_e^{\ell}$.

\end{itemize}
 
The data structure on Alice's side stores $\bar{E}_A$ and some related auxiliary data; the data structure on Bob's side stores $\bar{E}_B$ and some related auxiliary data. Recall that $\bar{E}_A$ and $\bar{E}_B$ are subtrees of the common canonical protocol tree $\mathcal{T}$, which is a binary tree of depth $n$, and we will always have $|\bar{E}_A|=O(n)$ and $|\bar{E}_B|=O(n)$. In the following, we explain the data structure on Alice's side. The one on Bob's side is similar. 

The only \emph{update} operation on the data structures will be to \emph{add an edge} to $\bar{E}_A$ and update the related auxiliary data. There are furthermore three query types, one related to each of the objectives listed above. We will provide an implementation that achieves an amortized time complexity of $\tilde{O}(1)$ per edge addition and a worst-case time complexity of $\tilde{O}(1)$ for each of the queries. It seems possible to de-amortize the time complexity of edge additions, but for sake of simplicity we do not do this here. Next, we state the implementation of each of these operations and explain how this implementation achieves the claimed $\tilde{O}(1)$ complexities. 

\subsubsection{The Binary Search Tree}\label{sec:binarysearchdatastructure}
The first query is concerned with how, at each point in time, Alice chooses the the next edge $e\in \bar{E}_A$ for which to check whether $P_e\subseteq\bar{E}_B$ or not. These edges $e$ form a binary search on $\bar{E}_A$. In particular, we would need the operation that receives an edge $e$ and outputs the two next edges $e'$ and $e''$ in $\bar{E}_A$ that are the binary search points after $e$. One edge, say $e'$, will be the next edge after $e$ that is used when we discard the edges outside $P_e$ and try to extend $P_e$ downwards; the other edge $e''$ will be the next edge that is used when we discard the edges on the subtree below $e$.
\begin{figure}[t]
	\centering
		\includegraphics[width=0.3\textwidth]{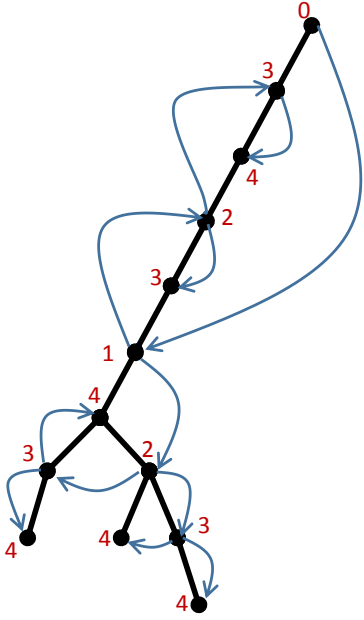}
	\caption{\small{A pictorial example of the semi-balanced binary search tree $\mathsf{BST}_A$ on top of $\bar{E}_A$. The black edges indicate those in $\bar{E}_A$ and the blue directed edges are downwards edges on the $\mathsf{BST}_A$. Note that for simplicity, we always root $\mathsf{BST}_A$ on the root of $\bar{E}_A$, which is the root of the binary protocol tree $\mathcal{T}$. The red numbers next to the nodes show their depth on $\mathsf{BST}_A$.}}
	\label{fig:BST}
\end{figure}
We store the binary search moves of Alice on $\bar{E}_A$ as a \emph{binary search tree} $\mathsf{BST}_A$. \Cref{fig:BST} shows an example. Note that edges can be uniquely identified with their lower end nodes. Each node of $\mathsf{BST}_A$ is a node at the lower end of an edge $e \in \bar{E}_A$ and we have the guarantee that $\mathsf{BST}_A$ is always semi-balanced, that is, for a node at the end of $e$ on $\mathsf{BST}_A$ and its two children, the subtrees of $\mathsf{BST}_A$ below these two children each have size at least $1/5$ of that below $e$. When we add an edge to $\bar{E}_A$, by default we add it below its $\bar{E}_{A}$-partent edge in $\mathsf{BST}_A$. Clearly such additions might lead to a violation of the mentioned semi-balance guarantee. As usual with data structures with amortized time complexities, we will preserve the semi-balance in a \emph{lazy} manner: for each node of $\mathsf{BST}_A$, we keep track of the size of the $\mathsf{BST}_A$-subtree below it. When we add an edge $e$ to $\bar{E}_{A}$, in $\mathsf{BST}_A$ we update the counts on all the $\log n$ nodes above (the lower end of) $e$ on $\mathsf{BST}_A$. If all nodes still have the semi-balance property, we are done. Otherwise, we pick the topmost node above $e$ for which the semi-balance is lost and we reconstruct the whole $\mathsf{BST}_A$ subtree below it such that everywhere below we have the stronger semi-balance guarantee of each side having size at least $1/3$.

\begin{lemma}\label{lem:BST-recomp}
Over any $N$ edge-additions to $\bar{E}_A$, the total time to recompute the binary search tree $\mathsf{BST}_A$ is $\tilde{O}(N)$.
\end{lemma}
\begin{proof}
Consider a subtree of $\mathsf{BST}_A$ that has a $\frac{1}{3}$-semi-balance and has size in range $[x, 1.1 x]$ for some value $x \in [1, O(n)]$. Reconstructing a sub-tree with size $\Theta(x)$ takes $\tilde{O}(x)$ time and when we reconstruct we generate a $\frac{1}{3}$-semi-balance. On the other hand, we reconstruct only when a $\frac{1}{5}$-semi-balance is broken, which can happen only after at least $\Theta(x)$ additional edges have been added. Over the $O(n)$ edge additions, there can be at most $\Theta(\frac{n}{x})$ reconstructions of subtrees with size in range $[x, 1.1 x]$. This is because, for each edge $e$ that gets added, there are at most $O(1)$ subtrees that include $e$ and have size in this range. This follows from the fact that because of the $\frac{1}{5}$-semi-balance, the size of the subtree below a node is at most a $\frac{4}{5}$ factor of that of its parent. Now, we know we spend a time of at most $\tilde{O}(x)$ for each reconstruction of a subtree with size in range $[x, 1.1 x]$ and there are in total at most $O(\frac{n}{x})$ such reconstructions. Therefore, the total time for reconstructions of subtrees with size in range $[x, 1.1 x]$ is $\tilde{O}(n)$. Now dividing the range $[1, O(n)]$ into $O(\log n)$ ranges $[1.1^{i}, 1.1^{i+1}]$, we get that overall the $O(n)$ edge additions, the total reconstruction time is $\tilde{O}(n)$.
\end{proof}

\subsubsection{Upward pointers} The second query type receives an edge $e\in \bar{E}_A$ and a length $\ell$ and finds the edge $e'$ that is $\ell$ levels above $e$ on the path $P_{e}$ connecting $e$ to the root of binary protocol tree $\mathcal{T}$. For this purpose, each edge $e$ keeps $\log n$ \emph{upward pointers}, one for each $i \in [0, \log n]$, which points to the edge $2^i$ levels up on $P_{e}$. Using these pointers, we can find $e'$ that is $\ell$ levels above $e$ in $O(\log n)$ steps. For instance, suppose that $\ell=11$. We find the edge $e_1$ that is $8$ levels above $e$, then the edge $e_2$ that is $2$ levels above $e_1$, and then the edge $e'$ that is $1$ level above $e_2$. 
\begin{corollary} Given an edge $e$ and a length $\ell$, we can find the edge $e'$ that is $\ell$ levels above $e$ on the path $P_{e}$ in time $O(\log n)$.
\end{corollary}

\begin{lemma} For a new edge added to $\bar{E}_A$, the upward pointers can be constructed in time $O(\log n)$.
\end{lemma} 
\begin{proof}
Suppose the $add$-$edge(e)$ operation is called. We denote the edge $e$ by $e_0$. Next we explain how to quickly find the edge edge $e_i$ which goes $2^i$ levels above $e$ for any $i\in [1, \log n]$ given that the edges $e_j$ for $j<i$ are already constructed. For this we simply follow the edge $e_{i-1}$ and then follow from there the edge going $2^{i-1}$ levels above the this edge edge $e_{i-1}$. Summing up these distances shows that this brings us in only two steps to the edge $e_{i}$ which can then be used to find the edge $e_{i+1}$ and so on. In this way we generate each upward pointer of the new edge in in $O(1)$ steps for a total of $O(\log n)$ steps.
\end{proof} 


\subsection{Hash Comparisons in Polylogarithmic Time Using Splittable Codes}\label{sec:outlinefasthashing}

The new search procedure given in \Cref{sec:outlinedoublebinarysearch} reduces a tree-intersection to $\tilde{O}(1)$ pairwise (hash) comparisons of paths which correspond to $\tilde{O}(n)$ long binary strings to test these strings for equality with a somewhat small failure probability. In this section we explain how one can perform such a hash any hashing quickly if paths are stored in appropriatly coded form, and how maintaining such an encoding can be integrated efficiently into the binary search data structure from \Cref{sec:binarysearchdatastructure}. 

We first describe a well-known connection between hashing and sampling error correcting codes that was also used in a similar, but much simpler, way in \cite{BN13} to speed up hash computations. 

\subsubsection{Fast Hashing by Subsampling Error Correcting Codes}

Suppose two parties, Alice and Bob, are each given two bit strings of length $n$, say $x_A,x_B$, and they want to determine whether $x_A = x_B$ using the minimum amount of communication and computation. It is easy to see that, if the parties are required to be always correct, this task cannot be performed using less than $n$ bits of communication and therefore also $O(n)$ time. However, if the parties only want to be correct with say probability $p$ the communication complexity can be improved drastically. In particular, each of the parties can independently sample a uniformly random $\Theta(\log n)$-bit seed $s$ to select a random hash function $h_s$ and send the other party both this seed $s$ and the hash value of its string, e.g., $h_{s_A}(x_A)$. The other party can then apply the same hash function on its string, e.g., compute $h_{s_A}(x_B)$, and compare the outcomes. For good families of hash functions with a seed length $|s| = \Theta(\log n)$ the probability of having a hash collision, that is, a matching hash of two non-equal strings, is at most $1/n$. Repeating this $i$ times with independent random seeds one can boost the failure probability to $1/n^i$ using only $\Theta(i \log n)$ rounds of communication, which is optimal. However, each hashing step still requires $O(n)$ computations and even if parties have constant time random access to the strings this computation time cannot be reduced. 

However, if both parties are given random access not to $x_A$ and $x_B$ but instead to an error correcting encoding $x'_A$ and $x'_B$ under the same error correcting code of distance, say, $1/8$ one can perform hashing much faster. The reason for this is that the distance property of the code guarantees that if $x'_A \neq x'_B$ then both strings differ in at least a one eighth fraction of the positions. Now one can for example use $i \log n$ bits to sample $i$ independent uniformly random positions in these strings and compare only these positions using $i \log n$ rounds of communication to exchange the positions and $i$ rounds of communication to the symbols of $x'_A$ and $x'_B$ on these positions. Since each such sample uncovers a difference between $x_A$ and $x_B$ with probability at least $1/8$ if such a difference exists this results in a $1/8^i$ failure probability. The following lemma gives a slight improvement over this by generating slightly dependent positions, e.g., using a random walk on a constant degree expander, which still result in essentially the same failure probability while requiring fewer bits of randomness to be sampled:

\begin{lemma}[\cite{goldreich2011sample}] \label{lem:expanderwalkpositions}
Given any two bit strings $x,y$ of length $n$ which differ in at least $n/8$ positions and given $C = \Omega(\log n)$ uniformly random bits sampled independently of $x,y$ one can, in $\tilde{O}(1)$ time compute $\Theta(C)$ positions such that the probability that $x$ and $y$ agree on these positions is at most $2^{-\Theta(C)}$. 
\end{lemma}

\begin{corollary}\label{lem:fastregularhashing}
Given random access to encodings of their strings $x_A$ and $x_B$ of length $O(n)$ and any $C = \Omega(\log n)$ two parties can compare equality of these strings using $O(C)$ communications and $\tilde{O}(C)$ time computations up to a failure probabily of $2^{-C}$, if the encoding is an error correcting code with constant distance.
\end{corollary}

\subsubsection{Using Splittable Codes for Fast Hashing}

We want to use an equivalent of \Cref{lem:fastregularhashing} to compare pathes. However, it will not be possible to maintain an encoding of every path. Instead we will maintain \emph{splittable encodings} of certain subintervals of these pathes whith which we can cover any path using with a small number of these intervals. The next lemma states that fash hashing can still be performed when one has access to such splittable encodings.

\begin{lemma}\label{lem:fastsplittablehashing}
Given random access to encodings of their strings $x_A$ and $x_B$ of length $O(n)$ and any $C = \Omega(k \log^3 n)$ two parties can compare equality of these strings using $O(C)$ communications and $\tilde{O}(C)$ time computations up to a failure probability of $2^{-C}$, if the encoding of a string $x$ corresponds to $k$ sub-intervals encoded with splittable codes with these sub-intervals covering the string $x$ completely. This remains true if the way $x_A$ and $x_B$ are covered are not aligned. 
\end{lemma}

We first define splittable codes:

\paragraph{Splittable-code}
The splittable encoding of a string $S$ of length $n$ for a offset $x$ is as follows: A collection of error correcting codewords of parts of $S$ are stored in $\log n$ \emph{levels}. For each level $i\in [1, \log n]$, for each  bit $j$ of $S$ such that $2^{i-1}|(j+x)$, we say bit $j$ is an \emph{$i^{th}$-level break point}. A substring of length $2^{i-1}$ between two consequent $i^{th}$-level break points is called an \emph{$i^{th}$-level substring}. We will store $\log n$ levels of encoding of $S$ as follows. For each level $i \in [1, \log n]$, we store a coded version of each of the $i^{th}$-level substrings that are completely inside $S$. This coding is performed with an efficient error-correcting code $\mathcal{C}_i$ with constant distance. Thus, for a given string $S$ with length $\ell$, we store at most $O(\ell \log n)$ bits and all the levels of encoding can be computed in $\tilde{O}(\ell)$ time. We store the levels of encoding in a tree rooted at the level $\log n$ substring, and for each code in each level, the code is stored in an array so that we can have query access to each of its positions in $O(1)$ time. An example of this tree structure is shown in \Cref{fig:splittable-code}. 

\begin{figure}[t]
	\centering
		\includegraphics[width=0.9\textwidth]{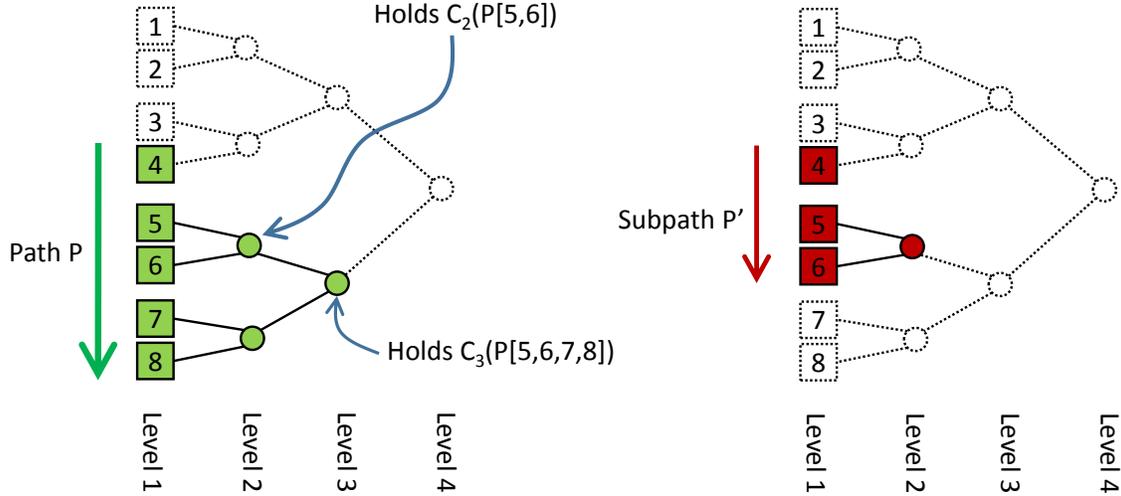}
	\caption{\small{A pictorial example of what is stored in the splittable code. The figure on the left is about what is stored while encoding a string $x$ with length $5$ and offset $3$. The splittable-code tree stores the levels of encoding, and the green nodes are those that the related substring is inside $x$ and thus their encoding will be stored. Each level-$i$ encoding node stores an error-correcting code $\mathcal{C}_i$ of the string on the leaves of the subtree below this node. The encoding stored by two example nodes on levels 2 and 3 are shown. The splittable-code of $x$ contains the splittable code of any subpath $x' \subseteq x$. An example is shown on the right side where the nodes of the encoding of a subpath $x'$ are indicated with red.}}
	\label{fig:splittable-code}
\end{figure}

The following observation captures an important property of a splittable-code that is also responsible for its naming:

\begin{observation}\label{obs:splitting}
For any string $x$ of length $n$ and any subinterval $x' = x[i,j]$ of $x$ the splittable encoding of a $x$ with offset $x$ contains the splittable encoding of $x'$ with offset $x+i-1$.
\end{observation}

It is also easy to see that such a sub-encoding can easily be identified and accessed. In particular we can identify the splittable-code data structure of $x'$ such that we have $O(\log n)$-time query access to each position of it. \Cref{fig:splittable-code} shows an example. To identify the splittable-code of $x'$, we walk over the the splittable-code tree of $x$ from the root downwards: for each node that its whole interval is in $x'$, this node and all of its descendant nodes are included. For any node that its interval is not fully contained in $x'$, we move to the children of it that their intervals has an overlap with $x'$ and we repeat the operation on those nodes. It is easy to see that during this walk, we will visit at most $O(\log n)$ nodes of the splittable-code tree. 

Next we prove \Cref{lem:fastsplittablehashing}:

\begin{proof}[Proof of \Cref{lem:fastsplittablehashing}]
Both parties have splittable encodings that cover their strings and thanks to \Cref{obs:splitting} one can safely assume that these splittable encodings actually partition their strings into at most $k$ pieces each, as any overlapping parts of a covering can be split off. In the first $O(k \log n)$ rounds of communication Alice and Bob exchange the cutting points of their partitioning and then split their splittable encodings to the common partitioning with the smallest number of cut points. This increases the number of cut partitions to at most $2k$. Each splittable encoding of a part of this partition can furthermore be cut into $\log n$ codewords, which due to the consistent offsets are aligned between Alice and Bob. An exaple is given in \Cref{fig:Path-Comp}. These codewords have the property that if the strings disagree in the interval covered by the codeword the corresponding codewords disagree in at least a constant fraction of their symbols. This allows us to subsample each of these $2k \log n$ codewords. In particular, we use \Cref{lem:expanderwalkpositions} to generate $C$ positions from each codeword (codewords shorter than $C$ are simply kept entirely) using the same randomness. Restricting the string comparison only to these $C$ positions in each codeword preserves any disagreements with probability at least $1 - 2^{\Theta(C)}$. In particular with this probability the string formed by concatenating the $2Ck \log n$ symbols on the selected positions from all codewords leads to a different outcome on both sides. The equality of these $\tilde{O}(C)$ short strings can now be easily tested using standard hashing which requires $O(C)$ bits of communication, and $\tilde{O}(1)$ time computations while not changing the failure probability by more than a constant. In total only $O(k \log n)$ and  $O(C)$ bits are communicated, the first one for communicating the initial splitting points the later for the randomness to select the positions, for the hash function seed and the final hash value. 
\end{proof}

\begin{figure}[t!]
\centering
\begin{minipage}{1.0\textwidth}
\centering
\includegraphics[width=0.85\textwidth]{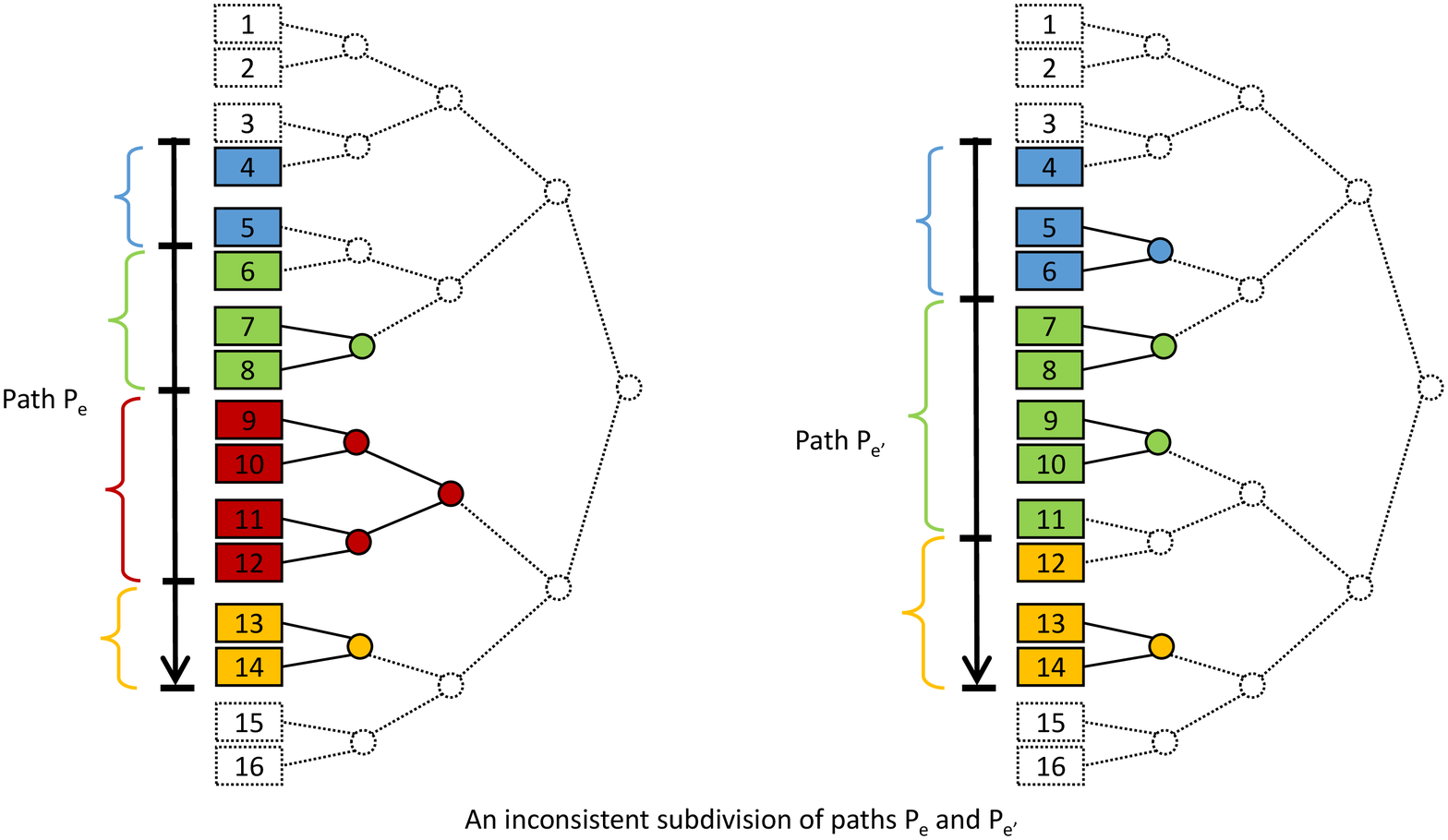}
\end{minipage}\\[0.7cm]
\begin{minipage}{1.0\textwidth}
\centering
\includegraphics[width=0.85\textwidth]{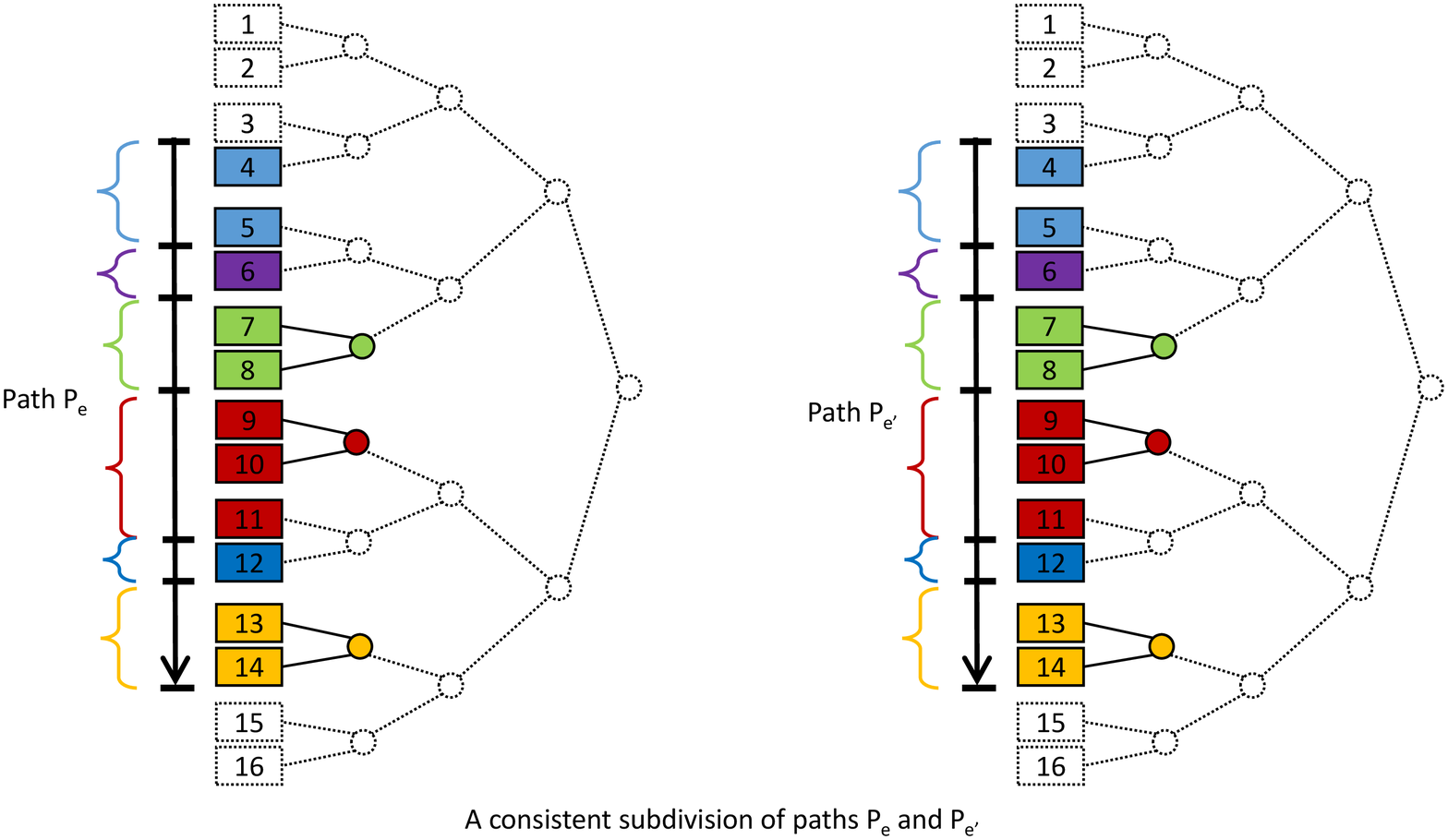}
\end{minipage}\\[0.7cm]

\caption{\small An example of an inconsistent subdivisions of paths $P_e$ and $P_{e'}$ that are made consistent by taking the union of cut points.}
\label{fig:Path-Comp}
\end{figure}

\subsubsection{Path Comparisons using Splittable Codes}\label{lem:path-comp-with-splits}

In this last section we explain how splittable codes and the hash comparison scheme of \Cref{lem:fastsplittablehashing} can be combined with the binary search tree data structure to obtain a fast tree-intersection algorithm. 

The most important question is which $\mathcal{T}$-paths one should store and maintain splittable encodings for in the data structure. Note that we cannot afford to store the splittable code of the whole path $P_{e}$ for each edge $e$ that gets added to $\bar{E}_A$. This is because constructing that would require time at least linear in the length of $P_{e}$ and it is possible that for many of the $O(n)$ edge-additions, this length is around $\Theta(n)$. Instead, the way we store $P_{e}$ will implicitly break it into $O(\log n)$ subpaths, using the structure of the binary search tree $\mathsf{BST}_A$:

For this, recall the binary search tree $\mathsf{BST}_A$ explained above which is a binary search defined over nodes of $\bar{E}_A$. Consider an edge $(u, v)$ in this binary search where $v$ is a child of $u$. Each of these nodes $u$ and $u$ is the lower end of an edge $\bar{E}_A$ and thus $u$ and $v$ are also on the binary protocol tree $\mathcal{T}$. We will store a splittable-code data structure of the $\mathcal{T}$-path $P_{uv}$ from $v$ to the lowest common ancestor of $u$ and $v$ on $\mathcal{T}$. We later explain how using these stored data, we can get ($\tilde{O}(1)$-time query access to) a splittable-code of each $\mathcal{T}$-path $P_e$ broken into at most $O(\log n)$ subpaths. Note that this is exactly what we used in the last paragraphs of \Cref{lem:path-comp-with-splits} when comparing two paths using their splittable codes.

We first show that for each node $v$ of the binary search tree $\mathsf{BST}_A$, the total length of the paths stored corresponding to edges in the $\mathsf{BST}_A$-subtree below $v$ is at most equal to the size of this subtree, up to logarithmic factors. Later we use this to argue that the total time to construct these splittable codes, over all the edge-additions, is $\tilde{O}(n)$.

\begin{lemma}\label{lem:total-length} Consider a node $v$ on the binary search tree $\mathsf{BST}_A$ and suppose that the $\mathsf{BST}_A$-subtree below $v$ is denoted by $\mathsf{BST}_A(v)$. Then, the total length of the $\mathcal{T}$-paths stored corresponding to edges in the $\mathsf{BST}_A(v)$ and thus also the total time for encoding them is $\tilde{O}(|\mathsf{BST}_A(v)|)$. 
\end{lemma}
\begin{proof}
For each node $u$ in $\mathsf{BST}_A(v)$, define the weight $\Phi(u)$ of $u$ to be the size of the subtree $\mathsf{BST}_A(u)$ that is below $u$, i.e., $\Phi(u)=|\mathsf{BST}_A(u)|$. It is easy to see that $$\sum_{u\in \mathsf{BST}_A(v)}\Phi(u) =O(\mathsf{BST}_A(v) \log n).$$ This is simply because the contribution of each node $w$ to this summation is one unit for each node $u$ that is above $w$ in the $\mathsf{BST}_{A}$ and since $\mathsf{BST}_{A}$ is a semi-balanced binary tree, there are at most $O(\log n)$ such nodes $u$. Now each node $u \in \mathsf{BST}_A(v)$ has at most $2$ children in $\mathsf{BST}_A(v)$ and for the $\mathsf{BST}_A$-edge to each of these children, we store the splittable code data structures of the related $\mathcal{T}$-path. We claim that each of these paths has length at most $O(\Phi(u))$. Once we have this claim proven, we immediately get that the total length of the $\mathcal{T}$-paths stored corresponding to edges in the $\mathsf{BST}_A(v)$ is at most $O(\mathsf{BST}_A(v) \log n)$. 

To prove the claim, consider a child $w$ of $u$ and let $P_{uw}$ be the $\mathcal{T}$-path to the lowest common ancestor of $u$ and $v$ in $\mathcal{T}$, for which we store the splittable code data structure. Also, let $u'$ be the $\mathsf{BST}_A$ parent of $u$. To show that $|P_{uw}|\leq \Phi(u')$, we simply 
show that each node of $P_{uw}$ is inside $\mathsf{BST}_A(u')$. For that, we revisit what happens when we pick $w$ as the next binary search node after $u$. There are two cases: If $w$ is in the $\mathcal{T}$-subtree below $u$, then $u$ itself is the lowest common ancestor of $u$ and $w$ and thus, the nodes of $\mathcal{T}$-path $P_{uw}$ are in $\mathsf{BST}_A(u)\subset\mathsf{BST}_A(u')$. The other case is if $w$ is not in the $\mathcal{T}$-subtree below $u$. In this case, what happened in the binary search is that we discarded the whole $\mathcal{T}$-subtree below $u$ and we picked some other node in the part of the $\bar{E}_A$ tree that is remaining active. Since at all times the active remaining part of $\bar{E}_A$ is a tree, and thus connected, the whole $\mathcal{T}$-path connecting $u$ to $w$ was present in the active remaining part of $\bar{E}_A$. Therefore, in particular the part of this path that connects $w$ to the lowest common ancestor of $u$ and $w$ in $\mathcal{T}$ was present which means that all the nodes of this path will be in $\mathsf{BST}_A(u')$, thus completing the proof. 
\end{proof}

Now note that for each path with length $\ell$, the time to generate the related splittable-code data structure is $\tilde{O}(\ell)$. We use this to show that the total time for generating splittable-codes over all the $O(n)$ edge-additions is $\tilde{O}(n)$.
\begin{lemma}
Over the $O(n)$ edge-additions to $\bar{E}_A$, the total time to compute the splittable-code data structures of all $\mathcal{T}$-paths corresponding to $\mathsf{BST}_A$-edges is $\tilde{O}(n)$.
\end{lemma}
\begin{proof}[Proof Sketch] 
Adding an edge to $\bar{E}_A$ and thus one to $\mathsf{BST}_A$ can possibly lead to a re-balancing (reconstruction) of $\mathsf{BST}_A$. That is, we might need to reconstruct the whole $\mathsf{BST}_A$ tree below a node $v$ from scratch, which means that we would need to also recompute the related splittable code data structures. From \Cref{lem:total-length}, we get that for a subtree with size in range $[x, 1.1x]$, the total length of the splittable code data structures in this subtree is $\tilde{O}(x)$. Morover, in the splittable-code data structure, we are using error-correcting codes with near-linear time encoding. Hence, the total time to generate the splittable-code data structures of this subtree is also in $\tilde{O}(x)$. Now similar to the proof of \Cref{lem:BST-recomp}, we know there are at most $O(\frac{n}{x})$ subtree reconstructions for subtrees with size in range $[x, 1.1]$. Again  similar to the proof of \Cref{lem:BST-recomp} by dividing the range $[1, O(n)]$ into $O(\log n)$ ranges $[1.1^{i}, 1.1^{i+1}]$, we get that overall the $O(n)$ edge additions, the total splittable-code construction time is $\tilde{O}(n)$.
\end{proof}

\paragraph{How to get an $O(\log n)$-subpath splittable code of a $\mathcal{T}$-path $P_e$} Consider a rooted $\mathcal{T}$-path $P_{e}\subseteq \bar{E}_A$ for which Alice wants to do a hash-check with a rooted $\mathcal{T}$-path $P_{e'}\subseteq \bar{E}_B$. 
Consider the node $v$ at the lower end of edge $e$ and let $v=v_0, v_1, v_2, v_3, \dots, v_{\ell}=r$ be the $\mathsf{BST}_A$-path that connects $v$ to the root of $\mathsf{BST}_A$, which is also the root of $\mathcal{T}$. Note that $\ell$ is at most the depth of $\mathsf{BST}_A$ which is $O(\log n)$. Now for each $\mathsf{BST}_A$-edge $(v_i, v_{i+1})$, we have stored a splittable coding data structure of the $\mathcal{T}$-path $P_{i, i+1}$ that connects $v_{i}$ to the lowest common ancestor of $v_{i}$ and $v_{i+1}$ in $\mathcal{T}$. Hence, if we put these paths $P_{i, i+1}$ together, we get a walk over $\mathcal{T}$ that covers goes from $v$ to the root $r$ and covers the $\mathcal{T}$-path $P_{e}$. To make this walk equal to $P_e$, we simply need to remove some part of each of the paths $P_{i, i+1}$. Thanks to the structure of the splittable codes, we can easily remove the splittable codes related to the redundant part of each of paths $P_{i, i+1}$ in $\tilde{O}(1)$ time. Thus, we now have the whole path $P_e$ broken into  $O(\log n)$ subpaths and for each of these subpaths, we have a splittable code data structure. This is exactly what we used in the two paragraphs at the end of \Cref{lem:path-comp-with-splits}, when comparing two paths using their splittable codes. }
\section{Proving the End Results}
In this small section, we explain how to combine the reduction technique of \Cref{sec:listdecodingreductions} with the list-decoders derived via boosting in \Cref{sec:boosting,sec:linearBoost} to prove the main end results of this paper, namely \Cref{thm:main14,thm:mainOthers} and \Cref{rmrk:deterministic}.

\begin{proof} [Proof of \Cref{thm:main14}]
To prove this theorem, we view the unique decoding coding scheme of Braverman and Rao\cite{BR11} as a list-decoding that tolerates error rate $1/4-\eps/2$. Thus, we get a deterministic list decodable coding scheme, with list size $1$, for any $O((\log \log \log n)^2)$-round protocol over a channel with alphabet size $O(1/\eps)$ and error-rate $1/4 -\eps/2$, round complexity $O((\log \log \log n)^2)$ and computational complexity $\tilde{O}(\log n)$. We then boost this list-decoder to a list-decoder for $n$-round protocols with communication complexity $N=O(n)$ using \Cref{lem:recBoosting}. As the result, we get a list-decoder for any $n$-round protocol that has round complexity $O(n)$, constant list size of $O(1/\eps^2)$, failure probability $2^{-\Theta(n)}$, and computational complexity $\tilde{O}(n)$. Then, we apply \Cref{thm:listdecodingreduction14}, which gets us to a unique decoder that tolerates error-rate $1/4-\eps$ and thus finishes the proof of \Cref{thm:main14}. 
\end{proof}

%

\begin{proof}[Proof of \Cref{thm:mainOthers}]
Braverman and Efremenko\cite{BE14} present a list-decodable coding scheme that tolerates error-rate $1/2-\eps/2$ and has linear communication complexity and exponential compuational complexity. By recursively boosting this list-decoder using  \Cref{lem:recBoosting}, we get a randomized list-decodable coding scheme that robustly simulates any $n$-round protocol  in $O(n)$ rounds tolerating error rate $1/2 - \eps$ with list size $O(1/\eps^2)$, computational complexity $\tilde{O}(n)$, and failure probability at most $2^{-\omega(n)}$. This already gives item (C) of \Cref{thm:mainOthers}. Combining this list decoder with \Cref{thm:listdecodingreduction27,thm:listdecodingreduction13} provides items (A) and (B) of \Cref{thm:mainOthers} respectively.

If one does not want to rely on the not-yet published result in \cite{BE14} one can also get essentially the same result by instead using the list decoder from \Cref{thm:logstar-boosted-simplified} which has an almost linear communication complexity of $N=n \cdot 2^{O(\log^* n \, \cdot \, \log{\log^* n})}$. This gives Item (C) directly and Items (A) and (B) follow again via Theorems \ref{thm:listdecodingreduction27} and \ref{thm:listdecodingreduction13}.
\end{proof}

\begin{proof}[Proof of \Cref{rmrk:deterministic}]
The proof is similar to proofs of \Cref{thm:main14,thm:mainOthers} with the exception that we use our non-uniform deterministic list-decoder boostings instead of the randomized ones. For the non-uniform deterministic variant of \Cref{thm:main14}, we boost the $1/4-\eps/2$ unique decoder of Braverman and Rao\cite{BR11} to get near-linear computational complexity using the non-uniform recurive boosting of \Cref{crl:recBoosting-deterministic}. Then, we combine it with the reduction in \Cref{thm:listdecodingreduction14}, which is a deterministic algorithm, gives us the non-uniform deterministic variant of \Cref{thm:main14}.
For the non-uniform deterministic variant of \Cref{thm:mainOthers}, we recurively boost the $1/2-\eps/2$ list decoder of Braverman and Efremenko via \Cref{crl:recBoosting-deterministic} to get near-linear computational complexity. This already gives us the non-uniform deterministic variant of item (C). Then, we use reductions \Cref{thm:listdecodingreduction27,thm:listdecodingreduction13} to get non-uniform deterministic variants of items (A) and (B). The alternative path, which does not rely on \cite{BE14}, would be to use \Cref{crl:logstar-boosting-deterministic} to get the non-uniform deterministic variant of item (C), and then again combine it with  reductions \Cref{thm:listdecodingreduction27,thm:listdecodingreduction13} to get non-uniform deterministic variants of items (A) and (B).
\end{proof}

\section*{Acknowledgements}
We thank Madhu Sudan for helpful discussions in the preliminary stages of this work that took place while the second author was a consulting researcher at Microsoft Research New England. 

\bibliographystyle{abbrv}
\bibliography{ref_eff}

\appendix
\shortOnly{
\section{Missing Parts of \Cref{sec:efficientadversaries}}\label{app:efficientadversaries}
\defSignature
\proofBoundedAdvlistdecodingreductionOneThird
\proofBoundedAdvlistdecodingreductionTwoSeventh
}

\shortOnly{
\section{Missing Parts of \Cref{sec:listdecodingreductions}}
\label{app:reduction}

}
\shortOnly{
\section{Missing Parts of \Cref{sec:boosting}}
\label{app:boosting}

}

\shortOnly{}
\section{One-Sided Unique Decoding}\label{app:oneway}

\subsection{Motivation for the One-Sided Unique Decoding Setting}

As described in \Cref{sec:interactivecodingsettings} the one-sided unique decoding setting only requires one a priori known party, say Alice, to decode. In this case one should think of Alice trying to compute something by (adaptively) querying Bob over an error prone channel. We provide two examples in which this is happens naturally:

Our first example might be a situation familiar to the reader
: Alice, a computer science researcher, tries to complete a proof but gets stuck often. She has a brilliant colleague, Bob, who Alice likes to consult and who knows the answers to all her questions. Alice wants to communicate with Bob such that she can finish her proof even if Bob and herself misunderstand each other a $\rho$ fraction of the time. Note that Bob might not know exactly what Alice proved in the end\footnote{He will however most likely have a good guess what Alice proved. For example, Bob might be able to give a short list of possible theorems that includes the one Alice proved. This is consistent with our observation that, even though the one-sided unique decoding setting does not make this an (explicit) requirement, Bob essentially must be able to list decode in the end in order to be helpful.}.

Our second, more technical, example involves a program or protocol that accesses data from a large remote database: Here, Alice is the program and Bob is the database. The program will send queries to the database, get results, and based on these results (adaptively) generate new queries. We want to guarantee that the program computes the correct output even if a $\rho$ fraction of the transmissions to or from the database are corrupted.  Again, we note that the only requirement is that the program produces the correct output, while it is not important whether the database knows in the end what the program computed or which of the queries it answered were (really) relevant for the computation. 

In all these scenarios it is an interesting question whether the fact that Bob is not interested in decoding allows the two parties to tolerate more errors. We show in \Cref{sec:reductions} that this is indeed the case.

%
%

\subsection{Impossibility Results for One-Sided Unique Decoding}\label{sec:onesidedLB}

In the remainder of this section we show that the $1/3 - \eps$ error rate achieved in item (B) of \Cref{thm:mainOthers} for a non-adaptive one-sided unique decoding scheme is best possible.

All the impossibility results proven in \cite{GHS13} rely on showing that even the much simpler exchange problem, in which both parties are given an $n$-bit input which they should transmit to the other party, is already impossible to solve under a certain error rate. In fact the impossibilities only use single-bit inputs. Looking at the exchange problem for a one-sided decoding setting, however, does not make sense as the exchange problem becomes equivalent to Bob sending data to Alice without Alice having to send anything. This is a classical one-way error-correction rather than an interactive coding. Instead, the simplest non-trivial protocol for one-sided unique decoding is the \emph{one-lookup problem}. In this problem, Alice gets an $n$-bit input $x \in \{0,1\}^n$ while Bob receives $2^n$ bits $y_1,\ldots,y_{2^n} \in \{0,1\}$. The \emph{$n$-bit one-lookup} protocol now consists of Alice sending her input $x$ to Bob and Bob replying with the bit $y_x$. In our motivating example this would correspond to Alice making a single (non-adaptive) look-up of a binary value in the database stored on Bob's side. 

We also remark that, in contrast to the proofs in \cite{GHS13}, proving meaningful impossibility results for error rates in the one-sided unique decoding setting is also not possible if one does not have (very moderate) restrictions on the alphabet size and round complexity. This is because in the one-sided unique decoding setting, it is always possible for Bob to encode all answers to all (adaptive sequences of) queries and send it to Alice. In our motivating example this would correspond to the database simply sending all its content to Alice without even trying to discriminate what information Alice requires. However, even for the simple one-lookup protocol above this takes exponential amount of communication. Taking this into account, our impossibility result therefore shows the next best statement: There is no non-trivial one-sided unique decoder tolerating an error rate of $1/3$. Differently speaking, any one-sided unique decoder that tolerates an error rate of $1/3$ needs to necessarily send exponential amount of data even if it just simulates the one-lookup protocol. 

\begin{theorem}\label{thm:onewayLBnonadaptive}
There is no non-adaptive one-sided unique decoder for the $n$-bit one-lookup protocol that tolerates an error rate of $1/3$, has a $o(n)$ alphabet size, and runs in $2^{o(n)}$ rounds while having a failure probability of $o(1)$.   
\end{theorem}
\begin{proof}
Suppose there is a $3N$ round protocol in which the adversary has an error budget of $N$. The adversary chooses both Bob's and Alice's inputs uniformly at random. Now, if Alice transmits for $N$ rounds or less, then the adversary simply corrupts all her transmissions. In this case, Bob has no information on Alice's input and cannot send an appropriate response. Since in $2^{o(n)}$ rounds with an $o(n)$ alphabet size Bob sends less than $2^{o(n)}$ bits to Alice the probability that she can guess the correct response is $o(1)$. Therefore, Alice needs to transmit for at least $N$ rounds. Furthermore, if Bob talks for $2N$ rounds or less then the adversary can corrupt half of Bob's transmissions and make it impossible for him to even transmit a single bit to Alice. Therefore, if the failure probability is $o(1)$, then Bob needs to talk more than $2N$ rounds. This shows that more than $3N$ rounds are necessary.
\end{proof}

Interestingly, the same $1/3$-impossibility also holds for adaptive coding schemes:						

\begin{theorem}\label{thm:onewayLBadaptive}
There is no adaptive one-sided unique decoder for the $n$-bit one-lookup protocol that tolerates an error rate of $1/3$, has an $o(n)$ alphabet size, and runs in $2^{o(n)}$ rounds while having a failure probability of $o(1)$.   
\end{theorem}
We defer the proof of \Cref{thm:onewayLBadaptive} to the full version of this paper and here present just a sketch of it:
\begin{proof}[Proof Sketch]
We explain the adversary's strategy:

Suppose we give Bob a random input. Suppose then we also give Alice a random input and with probability $1/2$ let her talk to the real Bob (until she has heard him N times) and then to a fake with different random input who has heard the first part of the conversation. And with probability $1/2$ we first let her talk to the fake Bob and then to the real one. 

There are two cases regarding which one of the following has probability at least $1/2$: the real Bob sends at least $2N$ times or he sends less than $2N$ times.

In the latter case where with probability at least $1/2$ Bob sends less than $2N$ times, the adversary does what is suggested above, that is it creates a fake Bob and lets him either talk first or second. This makes it impossible for Alice to determine which Bob is the right one and she therefore has a probability of at least $1/2$ to guess a wrong output if the adversary does not run over budget. The later is the case with probability at least $1/2$. This leads to a total failure probability of at least $1/4$ in this case. 

On the other hand, in the former case where with probability at least $1/2$ Bob sends at least $2N$ rounds then we simply create a fake Alice with random input and make her sound as above. Since Bob has no information about Alice's input he cannot transmit useful information and Alice will not be able to decode correctly with constant probability. 

\end{proof}

Lastly, we remark that the tolerable error rates for a one-sided decoder stay the same as for two-sided decoding for any of the other six settings discussed in \cite{GHS13}. In particular one-sided list decoding does not go beyond a tolerable error rate of $1/2$. Adding shared randomness also goes only to $1/2$ in the one-sided unique decoder setting (unless one puts together shared randomness and adaptivity in which case $2/3$ is the best achievable independently of whether one also allows one-sided unique decoding or list decoding for both sides). The upper bounds in \Cref{thm:main14,thm:mainOthers} and \cite{GHS13} together with the impossibility results in \Cref{thm:onewayLBadaptive} and \cite{GHS13} therefore settle the optimal tolerable error rate for all 16 possible interactive coding settings which are made by the Boolean attributes \{one-sided decoding / two-sided decoding\}, \{unique decoding / list decoding\}, \{adaptive / non-adaptive\}, and \{without shared randomness / with shared randomness\}. 

\end{document}